\documentclass[%
 aps,
 prb,%
 amsmath,amssymb,
 reprint,%
superscriptaddress,
]{revtex4-1}

\usepackage{graphicx}
\usepackage{dcolumn}
\usepackage{bm}
\usepackage{color}

\begin{document}
\newcommand{\micron}{$\mu$m}

\title{Equation of state of warm-dense boron nitride combining computation, modeling, and experiment}
\author{Shuai Zhang}
\email{zhang49@llnl.gov}
\author{Amy Lazicki}
\email{jenei2@llnl.gov}
\affiliation{Lawrence Livermore National Laboratory, Livermore, California 94550, USA}
\author{Burkhard Militzer}
\email{militzer@berkeley.edu}
\affiliation{Department of Earth and Planetary Science, University of California, Berkeley, California 94720, USA}
\affiliation{Department of Astronomy, University of California, Berkeley, California 94720, USA}
\author{Lin H. Yang}
\author{Kyle Caspersen}
\author{Jim A. Gaffney}
\author{Markus W. D\"ane}
\author{John E. Pask}
\affiliation{Lawrence Livermore National Laboratory, Livermore, California 94550, USA}
\author{Walter R. Johnson}
\affiliation{Department of Physics, 225 Nieuwland Science Hall, University of Notre Dame, Notre Dame, Indiana 46556, USA.}
\author{Abhiraj Sharma}
\author{Phanish Suryanarayana}
\affiliation{College of Engineering, Georgia Institute of Technology, Atlanta, GA 30332, USA}
\author{Duane D. Johnson}
\affiliation{Division of Materials Science \& Engineering, Ames Laboratory, Ames, Iowa 50011, USA}
\affiliation{Department of Materials Science \& Engineering, Iowa State University, Ames, Iowa 50011, USA}
\author{Andrey V. Smirnov}
\affiliation{Division of Materials Science \& Engineering, Ames Laboratory, Ames, Iowa 50011, USA}
\author{Philip A. Sterne}
\author{David Erskine}
\author{Richard A. London}
\author{Federica Coppari}
\author{Damian Swift}
\author{Joseph Nilsen}
\author{Art J. Nelson}
\author{Heather D. Whitley}
\email{whitley3@llnl.gov}
\affiliation{Lawrence Livermore National Laboratory, Livermore, California 94550, USA}

\date{\today}
\begin{abstract}
{
The equation of state (EOS) of materials at warm dense conditions poses significant challenges to both theory and experiment. We report a combined computational, modeling, and experimental investigation leveraging new theoretical and experimental capabilities to investigate warm-dense boron nitride (BN).
The simulation methodologies include path integral Monte Carlo (PIMC), several density functional theory (DFT) molecular dynamics methods [plane-wave pseudopotential, Fermi operator expansion (FOE), and spectral quadrature (SQ)], activity expansion (ACTEX), and all-electron Green's function Korringa-Kohn-Rostoker (MECCA), and compute the pressure and internal energy of BN over a broad range of densities and temperatures.
Our experiments were conducted at the Omega laser facility and  the Hugoniot response of BN to unprecedented pressures (1200--2650~GPa).
The EOSs computed using different methods cross validate one another in the warm-dense matter regime, and the experimental Hugoniot data are in good agreement with our theoretical predictions. 
By comparing the EOS results from different methods, we assess that the largest discrepancies between theoretical predictions are $\lesssim$4\% in pressure and $\lesssim$3\% in energy and occur at $10^6$~K, slightly below the peak compression that corresponds to the $K$-shell ionization regime.
At these conditions, we find remarkable consistency between the EOS from DFT calculations performed on different platforms and using different exchange-correlation functionals and those from PIMC using free-particle nodes. This provides strong evidence for the accuracy of both PIMC and DFT in the high-pressure, high-temperature regime.
Moreover, the recently developed SQ and FOE methods produce EOS data that have significantly smaller statistical error bars than PIMC, and so represent significant advances for efficient computation at high temperatures.
The shock Hugoniot predicted by PIMC, ACTEX, and MECCA shows a maximum compression ratio of 4.55$\pm$0.05 for an initial density of 2.26~g/cm$^3$, higher than the Thomas-Fermi predictions by about 5\%.
In addition, we construct new tabular EOS models that are consistent with the first-principles simulations and the experimental data. 
Our findings clarify the ionic and electronic structure of BN over a broad range of temperatures and densities and quantify their roles in the EOS and properties of this material.
The tabular models may be utilized for future simulations of laser-driven experiments that include BN as a candidate ablator material. (LLNL-JRNL-767019-DRAFT)}

\end{abstract}



\maketitle

\section{Introduction}\label{sec:introd}

The equation of state (EOS) of materials from the condensed matter to warm dense matter and the plasma regime plays an indispensable role in radiation hydrodynamic simulations~\cite{colvin_larsen_2013}, 
which are required for the design and analysis of inertial confinement fusion 
(ICF) and high energy density (HED) experiments. In laser-driven capsule experiments, ablator materials are important to implosion dynamics and performance. Currently, the most widely used ablator materials are plastics, such as 
polystyrene derivatives and glow-discharge polymer, high density carbon (HDC), and beryllium. Materials with higher density and tensile strength, such as boron (B) and its compounds, offer the potential for 
improvements in performance and additional nuclear diagnostics in exploding pusher platforms.~\cite{Ellison_2018,Zhang2018b}

At ambient conditions, BN exists in two stable, nearly degenerate phases: hexagonal BN (h-BN) and cubic BN (c-BN), similar to the graphite and diamond phases of its 
isoelectronic material, carbon (C).
Because of this similarity, BN is widely investigated for the synthesis of superhard materials and fabrication of thin films or heterostructures for various applications.~\cite{WANG2017MRP6}  
Nanostructured c-BN, whose hardness is almost twice that of bulk c-BN and close to that of diamond, has been synthesized at high-pressure and temperature conditions~\cite{Solozhenko2012am}.
Other applications for low-dimensional BN include nanoelectronic devices~\cite{WANG2017MRP6} and expanded h-BN for hydrogen storage~\cite{FU2017335CMS}.  It has also been demonstrated 
that the density and mechanical properties of BN can be tuned by constructing a mixture of its cubic and hexagonal phases.~\cite{DuFrane_2016}

There have been extensive theoretical and experimental studies on the structure~\cite{LONG2015638jac,Germaneau2013jpcm}, stability~\cite{BRITUN2000hpr,Sengupta2018prb,Taniguchi1997APL}, EOS~\cite{Aleksandrov1989,WANG2017276jpcs,Esler2010prl,Kawai2009jap,Hu2018jap,MarshLASL1980}, melting and phase diagram~\cite{Lee2016acsnano,Turkevich2002,Riedel1994,deKoker2012jpcm}, and mechanical~\cite{LeGodec2012jsm,Zhang2011jap,HAMDI20102785pbcm}, optical~\cite{Cunningham2018PRM,Attaccalite2011prb}, thermodynamic~\cite{WANG2017276jpcs,WANG20092082pla,Yang2009epjb,HAMDI20102785pbcm}, and transport~\cite{CHAKRABORTY201885,Jiang2018PRM} properties of BN and its polymorphs. The phase transformation of rhombohedral BN (r-BN) was found to be dependent on the pressure transmitting medium~\cite{Taniguchi1997APL}, and
the transition of h-BN into a wurtzite phase (w-BN) under plastic shear may be dramatically different from that under hydrostatic pressures~\cite{Ji2012PNAS,Levitas2004EPL}.
A large number of calculations using density functional theory (DFT)~\cite{ks1965,hk1964}, and quantum Monte Carlo (QMC) simulations~\cite{Esler2010prl,Ma2015prl,Atambo2015mre} have been performed on c-BN. Assisted by vibrational corrections, QMC results~\cite{Esler2010prl} successfully reproduce the volume changes and Raman frequency shifts measured by static high-pressure experiments.

Experimentally, 
the diamond anvil cell or multi-anvil apparatus have been used to obtain the EOS of h-BN up to $\sim$12~GPa and 1000~K~\cite{Lynch1966jcp,SOLOZHENKO19951ssc,FUCHIZAKI2008390ssc}, c-BN to 160~GPa and 3300~K~\cite{Datchi2007prb,Knittle1989Nature,Goncharov2007prb},
and of w-BN to 66 GPa~\cite{Solozhenko1998apl}.
Shock compression measurements for BN up to 300~GPa have been reported 
for various initial densities (1.81--3.48~g/cm$^3$)~\cite{Kawai2009jap,Hu2018jap,MarshLASL1980,Coleburn1968jcp}, porosity~\cite{MarshLASL1980}, and temperatures (293--713~K)~\cite{Coleburn1968jcp}.  
Because of the limited data available at extremely high pressure and temperature conditions, existing tabular EOS models have traditionally relied on simplified electronic structure theory, 
such as the Thomas-Fermi (TF) theory.  The goal of this work is to investigate the EOS of BN in the high-energy-density regime and provide new tabular models that are validated by first-principles simulations and experimental data.

In a recent study~\cite{Zhang2018b}, Zhang {\it et al.} computed the EOS of B based on first-principles quantum simulations over a wide range of temperatures and densities. The Hugoniot computed from those simulations  
shows excellent agreement with 
our experimental measurement on a planar laser shock platform. We have utilized the data to construct an EOS table (X52) for B. The work has also allowed us to study the performance of the polar direct-drive exploding pusher platform~\cite{Ellison_2018} and its sensitivity to the EOS. 

In this work, we combine extensive theoretical calculations to build tabular models for the EOS of BN, which we then validate in the warm dense matter regime via comparison to experimental measurements of the BN Hugoniot.  We 
also provide theoretical estimates of the uncertainty in the pressure and internal energy by comparing values from different simulation methods.  
Our theoretical methods include many-body path integral Monte Carlo (PIMC), several electronic structure theories based on pseudopotential DFT-molecular dynamics (DFT-MD), 
an activity expansion method, and 
an all-electron, Green's function Korringa-Kohn-Rostoker (KKR)
method. Our experiments consist of three measurements of the Hugoniot response of c-BN conducted at the Omega laser facility.

The paper is organized as follows: Sec.~\ref{sec:theorymethod}  introduces our simulation methods; 
Sec.~\ref{sec:expt} describes details of our shock experiments; 
Sec.~\ref{sec:eosmodel} introduces our EOS models;
Sec.~\ref{sec:results} compares and discusses our EOS and Hugoniot results from different theoretical methods and experiments and those between BN and C;
finally we conclude in Sec.~\ref{sec:conclusion}.

\section{First-principles simulation methods}\label{sec:theorymethod}

\begin{figure}
\centering\includegraphics[width=0.48\textwidth]{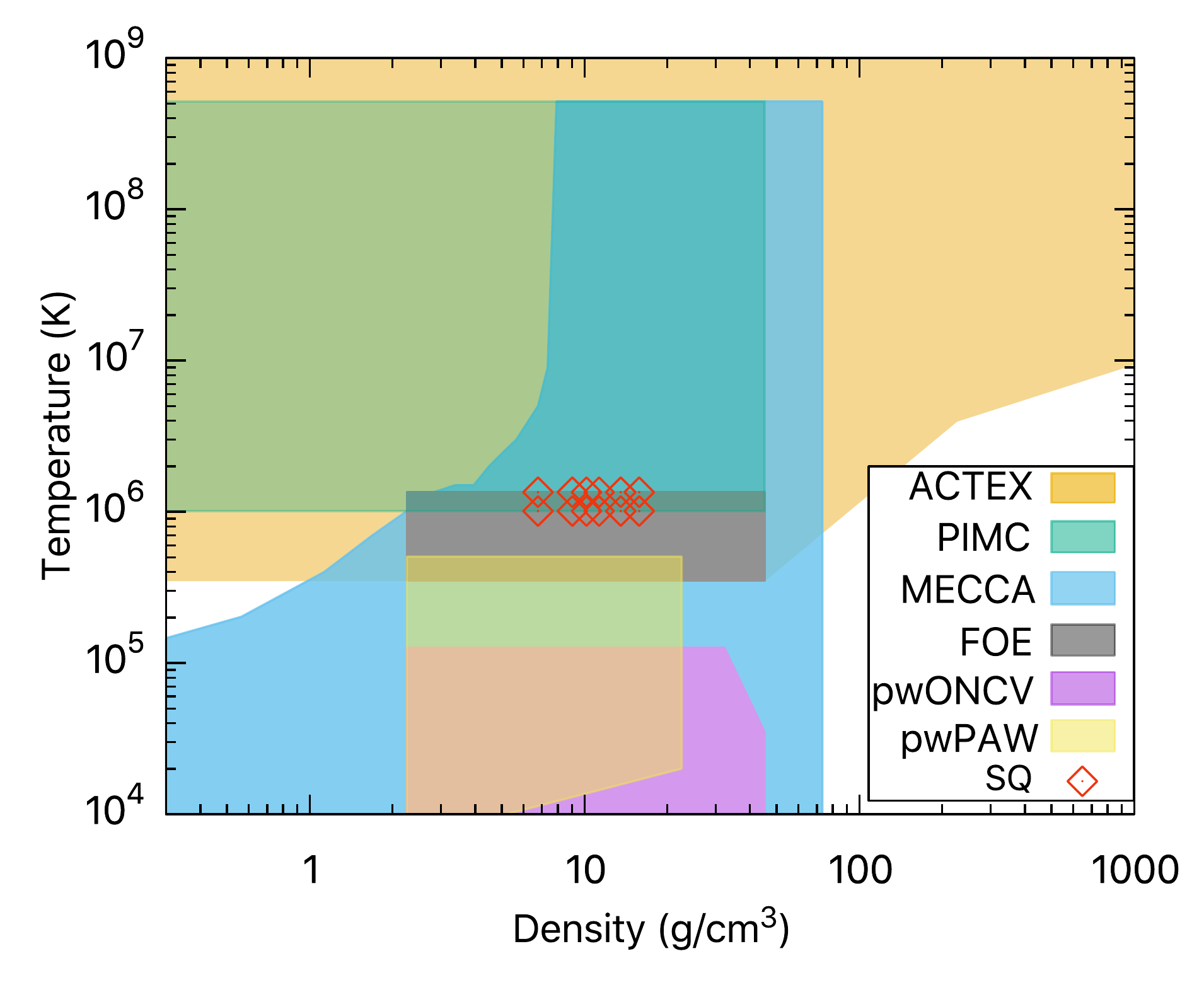}
\caption{\label{fig:theorymethods} Temperature-density  diagram showing the parameter regions where the methods in this article are used for calculating the EOS of BN.}
\end{figure}

In this section, we introduce the theoretical methods that are used in this work to compute the internal energies and pressures of BN across a wide range of temperatures and densities in order to 
provide simulation data for construction of new tabular EOS models for BN.  
The theoretical methods applied here include PIMC, the activity expansion method 
as implemented in the {\footnotesize ACTEX} code, and several methods that are based on DFT. The DFT methods include both methods that sample the ionic positions via molecular dynamics and 
average-atom methods where the ionic positions are static.  
Figure~\ref{fig:theorymethods} summarizes the temperature and density conditions at which each of the methods has been employed for calculations of BN in this study.
In the following, we briefly describe the fundamental assumptions associated with each technique and comment on its accuracy.  Additional details
can be found in the cited references.

\subsection{Path Integral Monte Carlo}\label{subsec:pimc}
PIMC is a quantum many-body method for materials simulations that is based on sampling the finite temperature density matrix derived from the full many-body Hamiltonian, $\mathcal{H}$.  
In PIMC, particles are treated as quantum paths 
that are cyclic in imaginary time [0,$\beta$=$1/k_\text{B}T$],
where $k_\text{B}$ is the Boltzmann constant.
Thermodynamic properties, such as the internal energy,
are obtained by 
\begin{equation}
\bar{O}=\frac{1}{Z}\int\int dRdR'\left<R\left|\hat{O}\right|R'\right>\varrho(R,R';\beta)
\end{equation} 
in coordinate representation. $Z=\int dR\left<R\left|\hat{O}\right|R\right>$ is the partition function. $\varrho(R,R';\beta)=\left<R|\exp(-\beta\mathcal{H})|R'\right>$ 
is the density matrix. 
Trotter's formula~\cite{trotter1959} can be used to break up $\varrho(R,R';\beta)$ into $M$ slices,
each corresponding to an imaginary time step  $\tau=\beta/M$.
The method becomes exact in the limit of $\tau\rightarrow 0$. 
Higher temperatures require fewer points, and convergence with respect to the imaginary time step must systematically be tested for each system studied. 
In practice, one starts with a solution of the two-body problem and only employs the PIMC method to sample higher-order correlations. This pair density matrix approach is described in Refs.~\onlinecite{pdm,Pollock1984}.

The application of PIMC to electronic structure calculations requires certain approximations due to the 
fermion sign problem. Fermionic symmetry requires that a negative sign 
arises from the anti-symmetrical wavefunction.
This leads to the nearly complete cancellation of positive and negative contributions to the fermionic density matrix, which makes a direct numerical evaluation impractical for more than a few particles.
The standard way to avoid this issue in PIMC simulations is to restrict the paths to the positive region of the trial density matrix, $\varrho_T$, by implementing the fixed-node approximation~\cite{Ceperley1991}. 
The condition $\varrho_T=0$ in 3$N$-dimensional space defines the nodal surface, where $N$ is the number of particles.
In high temperature simulations, $\varrho_T$ is chosen to be a Slater determinant of free-particle density matrices 
\begin{equation}
    \varrho^{[1]}(r_i,r_j;\beta)=\sum_k\exp(-\beta E_k)\Psi_k^*(r_i)\Psi_k(r_j), 
\end{equation}
where $\Psi_k^*(r)$
denotes a plane wave with energy $E_k$. The corresponding nodal surface is called free-particle nodes.  The assumption of free-particle nodes is appropriate
at high temperature.  
The PIMC method with free-particle nodes has been successfully developed and applied to hydrogen~\cite{PhysRevLett.73.2145,PhysRevLett.76.1240,PhysRevE.63.066404,PhysRevLett.87.275502,PhysRevLett.104.235003,PhysRevLett.85.1890,Hu2011,CTPP:CTPP2150390137,Militzer20062136}, helium~\cite{PhysRevLett.97.175501,Mi05}, and calculations of the EOS for a range of first-row elements~\cite{Driver2012,Driver2015Oxygen,Driver2016Nitrogen,Driver2015Neon,Zhang2018b} and compounds~\cite{Driver2012,Driver2017LiF,Zhang2017b,Zhang2018}.
Recent developments~\cite{Militzer2015,Zhang2017,Driver2018Al} have extended the applicability of PIMC to second-row elements at lower temperatures by appending localized orbitals to $\varrho^{[1]}$,
opening a possible route toward accurate quantum many-body simulations of heavier elements.

In this study, we apply PIMC for the simulations of BN with free-particle nodes using the {\footnotesize CUPID}  code~\cite{militzerphd}.
All electrons and nuclei are treated explicitly as quantum paths.
The Coulomb interactions are described via pair density
matrices~\cite{Na95,pdm}, which are evaluated in steps of $\tau=\frac{1}{512}$~Hartree$^{-1}$ (Ha$^{-1}$). The
nodal restriction is enforced in much smaller steps of $\frac{1}{8192} ~\text{Ha}^{-1}$.
The calculations are performed over a wide range of densities 0.23--45.16~g/cm$^3$, or 0.1- to 20-times the
ambient density $\rho_0\sim2.26$~g/cm$^3$ based on that of h-BN~\cite{hexBN}, and temperatures $10^6$--5$\times$10$^8$~K.
Each simulation cell consists of 24 atoms, which is comparable to our previous simulations for pure B~\cite{Zhang2018b}, nitrogen (N)~\cite{Driver2016Nitrogen}, and hydrocarbons~\cite{Zhang2017b,Zhang2018}. The cell size effects on the EOS are negligible at such high temperature conditions~\footnote{By comparing the EOS and the radial distribution function $g(r)$ obtained using 24-atom cells to those using 96-atom cells in our DFT-MD calculations, we find negligible differences at temperatures above $5\times10^4$ K. A comparison in $g(r)$ is shown in Fig.~\ref{fig:gr}.}.

\subsection{\label{sec:vasp}DFT-MD with plane-wave basis and projector augmented wave potentials}
DFT-MD is a widely used method for accurately simulating condensed matter systems at finite temperatures. In DFT-MD, 
the ions are classical particles, which move according to Newton's classical equations of motion.
The forces 
are computed 
by solving
the Kohn-Sham DFT equations for the electrons at each time step. The applicability and accuracy of DFT-MD for EOS calculations has been previously demonstrated for condensed phase materials in multiple studies (see Ref.~\onlinecite{Soderlind2018} as an example).
One difficulty lies in using this method for high temperatures, which is originated from significant thermal excitation of electrons and intractable computational cost.

Our DFT-MD simulations for BN are performed in two different ways. One way is by using the plane-wave basis and projector augmented wave (PAW) pseudopotentials~\cite{Blochl1994} (pwPAW), as implemented in the
Vienna \textit{Ab initio} Simulation Package
({\footnotesize VASP})~\cite{kresse96b} and 
used in our previous studies (e.g, Refs.~\onlinecite{Zhang2016b,Zhang2017,Zhang2017b,Zhang2018,Zhang2018b}).
Similar to our recent work on pure B~\cite{Zhang2018b},
we choose the hardest PAW potentials available in {\footnotesize VASP},
which freeze the 1s electrons in the core and have a core radius of 1.1 Bohr for both B and N. 
We choose the
Perdew-Burke-Ernzerhof (PBE)~\cite{Perdew96} functional 
for describing electronic exchange and correlation interactions, a large cutoff energy of 2000~eV for the plane-wave basis, and the
$\Gamma$ point to sample the Brillouin zone.  The simulations are carried out using a Nos\'{e} thermostat~\cite{Nose1984} to generate MD trajectories in the canonical ensemble.
The MD time step is chosen to ensure total energy conservation and takes on values of 0.05-0.55~fs in these calculations, with smaller values corresponding to higher temperatures.   
We typically run for 5000 steps at each density-temperature ($\rho-T$) condition, which is found to be sufficient for convergence of the computed energies and pressures.

To ensure consistency with the all-electron PIMC energies, our pwPAW energies from {\footnotesize VASP} reported in this study are shifted by 
-79.017~Ha/BN. This is determined 
with all-electron calculations for isolated B and N atoms
with {\footnotesize OPIUM}~\cite{opium} using the PBE functional.

Our pwPAW calculations are performed at 
temperatures between
6.7$\times$10$^3$~K and 5.05$\times 10^5$~K ($\sim$0.6--43.5~eV).
Due to limitations in applying the plane-wave expansion for orbitals
at low densities and limitations in the applicability of the pseudopotentials that freeze the 
1s$^2$ electrons in the core at high densities, we consider a smaller 
range of densities ($\rho_0$ up to 10$\times\rho_0$) than that was examined via PIMC simulations.  
These conditions are relevant to shock-compression experiments
and span the range in which Kohn-Sham DFT-MD simulations are feasible by conventional wavefunction based approaches.
We performed calculations with both 24-atom and 96-atom cells to minimize the finite-size errors.

\subsection{\label{sec:FOE}DFT-MD with optimized norm-conserving Vanderbilt pseudopotentials and Fermi-operator expansion}

As a check on the pwPAW calculations for the majority of the DFT-MD simulations and to enable extension of our DFT-MD calculations to higher density,
we perform a separate set of DFT-MD simulations by utilizing optimized norm-conserving Vanderbilt (ONCV)~\cite{oncv13,oncv13e} pseudopotentials---a plane-wave method (pwONCV) at low temperatures and a Fermi operator expansion method (FOE) at high temperatures---in order to verify our pwPAW calculations and expand the range of applicability of Kohn-Sham DFT to higher temperatures.
Detailed information about the ONCV pseudopotentials is described in Appendix A.

The pwONCV calculations at low temperature ($< 1.3\times 10^5$~K) are similar to those using pwPAW. We applied a preconditioned conjugate gradient method~\cite{Yang2007} to fully relax the 
electronic wavefunctions at each time step.  An efficient fast Fourier transform (FFT)
algorithm was used for the conversion of the wave functions between real and reciprocal spaces.
Each simulation is performed either with frozen 1s$^2$ core pseudopotentials (for $\rho\lesssim10\times\rho_0$) or with all-electron pseudopotentials (for $\rho>10\times\rho_0$), $NVT$ ensemble with over 5000 steps, time-step of 0.2 fs, and on 128-atom supercells. 



At temperatures greater than $3.5\times 10^5$~K, 
$K$-shell ionization becomes significant~\cite{Zhang2018b}.
We use all-electron ONCV potentials and FOE~\cite{Goedecker99,Bowler2012}, 
which takes advantage of the smooth Fermi-Dirac function at high temperature by approximating the function with polynomial expansion, to conduct Kohn-Sham DFT calculations.  
In the subspace-projected Hamiltonian approach, we adopted the Chebyshev filtered subspace iteration approach~\cite{Saad92}.
As the ground-state electron density
depends solely on the occupied eigenspace,
the technique exploits the fast 
growth property of Chebyshev polynomial to magnify the relevant spectrum, thereby providing an efficient approach for the solution of the Kohn-Sham eigenvalue problem.  
The matrix-vector 
multiplications in the Chebyshev filtering procedure are performed on the FFT grids in Fourier space and only considered if the vector has a non-zero value in the matrix. 

Three steps are involved in this method: (i) a Chebyshev filter to construct a subspace which is an 
approximation to the temperature-smearing occupied eigenspace in a given self-consistent iteration; (ii) FFT mesh to span the Chebyshev filtered subspace from real-space to Fourier space; 
(iii) FOE in terms of the subspace-projected Hamiltonian represented in the plane-wave basis to compute relevant quantities like the density matrix, electron density and band energy. The accuracy of the Chebychev polynomial expansion~\cite{Goedecker94,Goedecker95} depends on the electron temperature $T_\text{e}$, 
and the width of the eigenspectrum $\Delta E_\text{e}$. In particular, the degree of polynomial required to achieve the desired accuracy in the approximation~\cite{Goedecker94} of the Fermi-Dirac distribution is $\mathcal{O}(\Delta E_\text{e}/k_BT_\text{e})$.
A more accurate estimate that takes into account the location of the Fermi level can be found in Ref.~\onlinecite{suryanarayana2013spectral}.
Chebychev polynomial orders of 40--60 and localization radii ranging from 1.056 to 2.88  Bohr were used in the FOE method.

To achieve the same level of accuracy as the plane-wave approach, our
high-$T$ FOE simulations use PBE exchange-correlation functional and the same FFT meshes as the pwONCV method (real-space grid spacing ranges from 0.066 to 0.18 Bohr).  The $NVT$ simulations were carried out using 32-atom supercells. Each simulation involves 3000--6000 steps (0.05--0.1~fs/step) to ensure sufficient statistics.


\subsection{DFT-MD using spectral quadrature}

The spectral quadrature (SQ) method~\cite{suryanarayana2013spectral} is a density matrix based $\mathcal{O}(N)$ method for the solution of the Kohn-Sham equations that is particularly well suited for calculations at high temperature. 
In the SQ method, all quantities of interest, such as energies, forces, and pressures, are expressed as bilinear forms or sums of bilinear forms which are then approximated by quadrature rules that remain spatially localized by exploiting the locality of electronic interactions in real space~\cite{prodan2005nearsightedness}, i.e., the exponential decay of the density matrix at finite temperature~\cite{goedecker1998decay,ismail1999locality,benzi2013decay,suryanarayana2017nearsightedness}.
In the absence of truncation, the method becomes mathematically equivalent to the recursion method~\cite{HAYDOCK:1972p4639,HAYDOCK:1975p4638} with the choice of Gauss quadrature, while for Clenshaw-Curtis quadrature, the FOE~\cite{goedecker1994efficient,goedecker1995tight} in Chebyshev polynomials is recovered.
Being formulated in terms of the finite-temperature density matrix, the method is applicable to metallic and insulating systems alike, with increasing efficiency at higher temperature as the Fermi operator becomes smoother and density matrix becomes more localized~\cite{pratapa2016spectral,surprapa2018}.
$\mathcal{O}(N)$ scaling is obtained by exploiting the locality of the density matrix at finite temperature, while the exact diagonalization limit is obtained to desired accuracy with increasing quadrature order and localization radius. Convergence to standard $\mathcal{O}(N^3)$ planewave results, for metallic and insulating systems alike, is readily obtained \cite{pratapa2016spectral,surprapa2018}.

While mathematically equivalent to classical FOE methods for a particular choice of quadrature, the more general SQ formulation affords a number of advantages in practice~\cite{pratapa2016spectral,surprapa2018}. These include: 
(1) The method is expected to be more robust since it explicitly accounts for the effect of truncation on the Chebyshev expansion.
(2) The method computes only the elements of density matrix needed to evaluate quantities of interest---e.g., only diagonal elements to obtain densities and energies---rather than computing the full density matrix (to specified threshold) as in FOE methods.
(3) The method computes the Fermi energy without storage or recomputation of Chebyshev matrices as required in FOE methods.
(4) The method admits a decomposition of the global Hamiltonian into local sub-Hamiltonians in real space, reducing key computations to local sub-Hamiltonian matrix-vector multiplies rather than global full-Hamiltonian matrix-matrix multiplies as in FOE methods. Since the associated local multiplies are small (according to the decay of the density matrix) and independent of one another, the method is particularly well suited to massively parallel implementation; whereas the global sparse matrix-matrix multiplies required in FOE methods pose significant challenges for parallel implementation~\cite{Bowler2012}.

In the present work, we employ the massively parallel {\footnotesize SQDFT} code~\cite{surprapa2018} for high-temperature Kohn-Sham calculations. {\footnotesize SQDFT} implements the SQ method in real space using a high-order finite difference discretization wherein sub-Hamiltonians are computed and applied for each finite-difference grid point. For efficient MD simulations, Gauss quadrature is employed for the calculation of density and energy in each SCF iteration whereas Clenshaw-Curtis quadrature is employed for the calculation of  atomic forces and pressure~\cite{pratapa2016spectral}.
While applicable at any temperature in principle, the present implementation is most advantageous at temperatures in excess of $\sim10^5$~K,
where the Fermi operator becomes sufficiently smooth and density matrix sufficiently localized to reduce wall times below those attainable by standard $\mathcal{O}(N^3)$ scaling methods for the system sizes considered here;
though avenues exist to reduce this temperature substantially~\cite{xuSP2018}.

Simulations were carried out for a series of 32-atom BN unit cells at densities from 
6.77--13.55~g/cm$^3$ and temperatures from 1010479--1347305~K. All-electron ONCV~\cite{oncv13} pseudopotentials were employed for B and N with cutoff radii of 0.60 and 0.65~Bohr, respectively. Exchange and correlation were modeled in the local density approximation (LDA) as parametrized by Perdew and Zunger~\cite{Perdew81}. $NVT$ simulations were carried out using a Nos\'e-Hoover thermostat~\cite{Nose1984,Hoover1985} with $\sim$500 steps for equilibration followed by $\sim$3000--5000 steps for production (with time steps of 0.035--0.04 fs).
A finite difference grid spacing of $\sim$0.1 Bohr (commensurate with unit cell dimensions), Gauss and Clenshaw-Curtis quadrature orders of 50 and 76, respectively, and localization radius of 1.3 Bohr were employed in the SQ calculations to obtain energies to 0.02\% and pressures to 0.2\% 
(discretization error)
or less.

\subsection{All-electron, Green's function Korringa-Kohn-Rostoker}

In addition, we applied an all-electron, Green's function KKR electronic-structure method (based on Kohn-Sham DFT) implemented within a scalar-relativistic approximation, i.e., spin-orbit is ignored beyond the core electrons. We use the Multiple-scattering Electronic-structure Calculation for Complex Applications ({\footnotesize MECCA}) code, a $k$-space KKR code.~\cite{MECCA-2015}
More technical details on high energy density applications using {\footnotesize MECCA} and the advantages using a Green function method can be found in reference~\cite{WILSON201161}. 
{\footnotesize MECCA} is applicable to the whole pressure and temperature range of interest in this paper, beyond that available from pseudopotential methods. However, as presently implemented, {\footnotesize MECCA} is a static DFT code that does not sample the ionic degrees explicitly, i.e.,  vibrational energies and corresponding entropy contributions cannot be obtained. 
As such, one must add these either from another calculation or some analytic model. Here, we apply the ideal-gas correction to the {\footnotesize MECCA} results to provide the most consistent comparisons with the other methods.  This approach was used recently to address, for example, the principal Hugoniot curves for Be in a review of EOS models for ICF materials.\cite{Gaffney-etal-2018} 

For current results, we used the atomic sphere approximation with periodic boundary conditions to incorporate interstitial electron contributions to Coulomb energy from all atomic Voronoi polyhedra. 
The KKR spherical-harmonic local basis included $L_\text{max}=2$, i.e., $s$, $p$, and $d$ symmetries within the multiple-scattering contributions, and $L$'s up to 200 are included automatically until the free-electron Bessel functions contribute zero to the single-site wavefunction normalizations.  The Green's functions are integrated via complex-energy contours 
enclosing a subset of Matsubara poles at finite temperatures, as well as taking advantage of analytic continuation to decrease dramatically solution  times.~\cite{KKR-contour-1985, WILSON201161}
Various  DFT exchange-correlation functionals are included through use of the {\tt libXC} library.~\cite{libXC} 
In this work we used the LDA functional of Vosko, Wilk, and Nusair.~\cite{doi:10.1139/p80-159}
Brillouin zone integrations for self-consistent charge iterations were performed with a 16$\times$16$\times$16 Monkhorst-Pack~\cite{Monkhorst1977} $k$-point mesh along the complex-energy contour
for energies with an imaginary part smaller than 0.25~Rydberg, and a 10$\times$10$\times$10 $k$-point mesh otherwise.
A denser mesh was used for the physical density of states calculated along the real-energy axes when needed.


Even though BN occurs in many phases near ambient conditions, for simplicity we chose to use a dense packed but cubic structure, the B2 phase (CsCl prototype) for all {\footnotesize MECCA} calculation to cover the broad range of pressures and temperatures.

\subsection{Activity expansion}
Activity expansion calculations of the EOS are performed using the {\footnotesize ACTEX} code, which is based on an expansion of the plasma grand partition function in powers of the constituent 
particle activities (fugacities)~\cite{rogers73,rogers74}.  
The present calculations are similar to those used in previous work~\cite{Zhang2018b} and include interaction terms beyond the Debye-H\"uckel, electron-ion bound states and ion-core plasma polarization terms, along with relativistic and quantum 
corrections~\cite{rogers79,rogers81}. EOS data generated with the {\footnotesize ACTEX} code, as well as OPAL opacity tables which use the state populations computed from {\footnotesize ACTEX},
have been extensively checked by comparison with astronomical observations~\cite{rogers94} and with laser-driven experiments~\cite{rogers97}.
\par

As with previous studies~\cite{Zhang2018b}, we cut off {\footnotesize ACTEX} calculations at temperatures below the point where many-body terms become comparable to the leading-order Saha term ($T>5.8\times10^5$~K). This ensures that the activity expansion method is valid while allowing investigation of the predicted peak compression on the Hugoniot.

\section{Shock Hugoniot experiment}\label{sec:expt}

\begin{figure*}
\centering\includegraphics[width=0.95\textwidth]{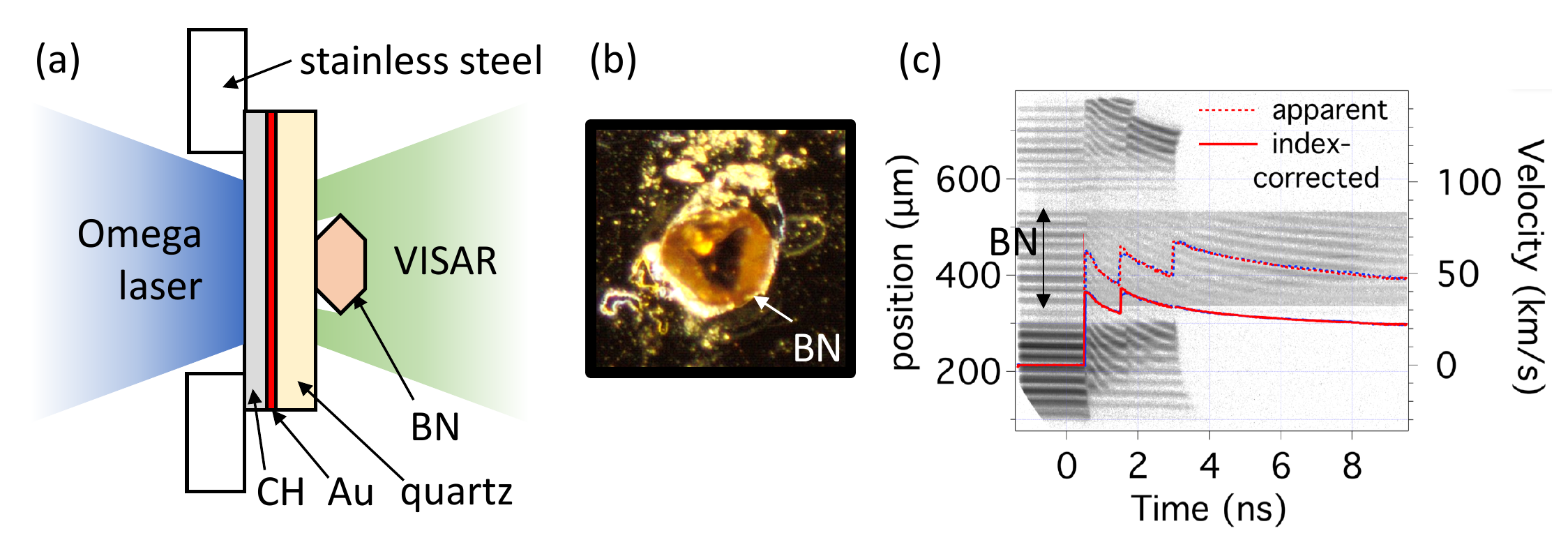}
\caption{\label{fig:expt} (a) experimental configuration (not drawn to scale), (b) image of a typical c-BN crystal glued to the quartz plate, viewed from the perspective of the VISAR diagnostic and (c) image of the VISAR data from shot 75265, with the analyzed velocities shown as red and blue traces (corresponding the two interferometer legs). The dashed traces are the apparent velocities and the solid traces are corrected for the index of refraction in quartz and cBN.}
\end{figure*}

Experiments to constrain the EOS of BN were performed at the Omega laser facility at the Laboratory for Laser Energetics in Rochester, NY.
Samples were c-BN crystals of greater than 99\% purity (by weight) and density of 3.45($\pm$0.03) g/cm$^3$, obtained from Saint-Gobain Ceramic Materials. Pale amber-colored $\{111\}$ and $\{\bar{1}\bar{1}\bar{1}\}$-oriented (identified by their morphology) optically transparent single crystals were characterized using x-ray photoelectron spectroscopy (XPS) and Raman spectroscopy as in~\cite{FENG2013817}.
XPS analysis was performed with a PHI Quantum 2000 system, using focused (1$\times$1 mm) monochromatic Al $K\alpha$ x-rays (1486.3 eV). XPS revealed a large amount of C, O and Si contamination, but a 60 second 3 kV Ar ion beam sputter (estimated to remove about 2-5~nm from the surface), dropped the concentration of contaminants by nearly 50\%, indicating that these form primarily a surface contamination (a $<1 \mu$m contaminated surface layer will have no effect on our measurement).
After etching, XPS identified a B:N ratio of 1.08:1. Room temperature Raman spectroscopy at 514.5~nm showed the TO and LO phonons of c-BN at 1057.7 and 1309.1~cm$^{-1}$, with no sign of the defect bands observed for amber crystals in Ref.~\onlinecite{FENG2013817}, indicating a high bulk purity. An extremely weak peak at 1122.3~cm$^{-1}$ suggests a negligible contamination of B$_4$C.

Crystals with parallel facets separated by $\sim$150~$\mu$m and lateral dimensions of 150-250~$\mu$m were affixed to $\sim$90~$\mu$m-thick $z$-cut $\alpha$-quartz (density of 2.65~g/cm$^3$) windows with micron-scale layers of epoxy. A 3-$\mu$m thick layer of Au was deposited on the other side of the quartz window, to absorb ablation plasma x-rays and reduce x-ray preheat of the BN samples to negligible levels, and a $\sim$25~$\mu$m-thick layer of plastic was deposited onto the Au to form the laser ablator (Fig.~\ref{fig:expt}(a)).

Samples were ablated directly using 12 beams at of the Omega laser with a 1-ns top-hat pulse shape and distributed phase plates forming a 800~$\mu$m spot size. Laser energies were tuned to drive the target at intensities ranging from $1.8\times10^{14}$ to $5\times10^{14}$~TW/cm$^{2}$.

A reflecting shock wave could be tracked continuously as it propagated through the quartz and c-BN samples, using a line-imaging velocimeter (VISAR: Velocity Interferometer System for Any Reflector)~\cite{Celliers_2004_VISAR}. The {\it in-situ} apparent velocities are corrected for the index of refraction of the quartz (1.54687)~\cite{GHOSH199995} and c-BN (2.126)~\cite{Gielisse1967} at 532~nm, which is the wavelength of the VISAR probe laser.

The shock velocities in the quartz and c-BN at the interface between the two are used in the impedance-matching technique, to determine the EOS data point for c-BN. Because of a finite glue bond thickness between the two materials, the shock velocity in the c-BN must be extrapolated to the quartz surface.
The quartz Hugoniot standard is taken from~\cite{Brygoo2015} and the reshock model from~\cite{Knudson2013}. The shock impedance in cBN at these conditions is higher than quartz, but sufficiently close that the accuracy of the off-Hugoniot quartz model has a small effect on the result (differs by $\sim$1\% from the result obtained by simply assuming a reflected Hugoniot for the reshock state).

\begin{table}
\centering
\begin{tabular}{c|c|c|c|c|c}
\hline  
& Quartz & \multicolumn{4}{c}{BN} \\
\hline  
& $U_s$ & $U_s$ & $U_p$ & $P$ & $\rho$ \\
& (km/s)    &     (km/s)  & (km/s)   & (GPa)  & (g/cm$^3$) \\
\hline 
75265 & 31.27(0.47) & 31.95(0.29) & 18.97(0.47) & 2091(53) & 8.49(0.34) \\
75263 & 34.99(0.34) & 35.04(0.31) & 21.87(0.37) & 2643(48) & 9.18(0.30) \\
75264 & 24.51(0.61) & 25.29(0.35) & 13.92(0.58) & 1214(52) & 7.67(0.44) \\
\hline 
\end{tabular} 
\caption{Measured quartz and c-BN shock velocities ($U_s$) and analyzed c-BN particle velocity ($U_p$), pressure ($P$) and density ($\rho$).}
\label{tab:expt}
\end{table}

The results of these measurements are recorded in Table~\ref{tab:expt}. Factors contributing to the uncertainty in the Omega measurements include: uncertainty in the quartz and c-BN wave velocities, uncertainty in the extrapolation of the c-BN velocity across the epoxy layer, uncertainty in the initial density of c-BN, and systematic uncertainty in the quartz standard EOS. Uncertainty in the c-BN index of refraction is not quantified so is not included in the error bar.

\section{Construction of EOS models for BN}\label{sec:eosmodel}
Before describing the results of the first principles simulations and experiments in detail, we describe the new EOS models and make comparisons to a subset of the calculations.  
We construct new EOS tables (X2151 and X2152) for BN under the QEOS framework~\cite{leos1qeos,leos2}. QEOS is a self-contained quasi-single-phase set of thermodynamic models that are widely applicable and guarantee the correct physical limits at both high/low temperature and high/low density.  The standard QEOS model based on  
TF theory also guarantees thermodynamic consistency.   
In our QEOS framework, we decompose the EOS into separate contributions corresponding to the $T=0$ cold curve, the 
ion thermal term that describes contributions to the EOS from the ionic degrees of freedom, and the electron thermal term that describes the contributions to the EOS from 
thermal distribution of the electrons.  
The cold curve is generally taken from experimental data
static DFT calculations, while the electron thermal term is generated using fast electronic structure methods, namely, 
TF theory and DFT calculations for the average atom-in-jellium model (Purgatorio) described in Appendix B.
The ion thermal term is often derived using a form proposed by Cowan~\cite{leos1qeos,leos2} and can be modified to fit both experimental data and data from many-body calculations.  In condensed phases (at high densities and low temperatures), the EOS, and hence the shock response 
of materials, is dominated by the cold curve, whereas the ion thermal term dominates the EOS through much of the high-velocity shock regime that is currently accessible in planar experiments at  
Omega and the National Ignition Facility.  The behavior of the EOS and the Hugoniot near peak compression, on the other hand, is mostly dominated by the electron thermal term.  The Hugoniot response that a model predicts  
near peak compression is therefore determined mostly by the underlying electron thermal model, and thus notable differences are seen between TF-based QEOS models and Purgatorio-based QEOS models.

The QEOS framework was chosen due to the lack of data necessary to constrain a more complicated multi-phase EOS 
representation and because the focus of the current study is in the liquid/plasma region relevant to high velocity, laser-driven shocks.
Both X2151 and X2152 tables have reasonably similar parameterization except for the 
electron-thermal model. At the time when the X2151 table was constructed there was only a Purgatorio~\footnote{We performed Purgatorio calculations for boron and for nitrogen in order to generate the electron thermal term of the EOS, which is used for constructing EOS tables based on the QEOS model.
Our Purgatorio calculations use the Coulomb potential and Hedin-Lundqvist~\cite{HedinLDA1971} form of exchange-correlation functional under LDA.}
electron-thermal model for B, therefore the full electron-thermal model for BN is a mixture of a  Purgatorio electron-thermal model for B and a TF electron-thermal model for N. Once a N Purgation electron-thermal model 
became available, the X2152 table was constructed, where the hybrid TF-Purgatorio electron-thermal model from X2151 was exchanged with a fully Purgatorio electron-thermal model (some adjustments to other EOS parameters were needed to 
improve the fit for X2152). Therefore, examining the L2150 (legacy TF EOS), X2151, and X2152 gives a demonstration of how the Hugoniot varies from a fully
mean-field TF description of ionization, to a hybrid treatment, to a fully quantum atom-in-jellium description.

\begin{table}
\centering
\begin{tabular}{lcl}
\hline 
 &  & Note \\ 
\hline 
$\rho_0$ & 2.258~g/cm$^3$ & reference density \\ 
$T_0$ & 295~K & reference temperature\\
$K_\text{h-BN}$ & 37~GPa & bulk modulus \\
$K_\text{c-BN}$ & 369~GPa & bulk modulus \\
$E_\text{coh}$ & 9$\times$10$^{10}$~erg/cm$^3$ & cohesive energy \\
$T_\text{m}^0$ & 2200~K & melt temperature @ 1~bar \\
$\Theta_\text{D}^0$ & 1675~K & Debye temperature @ $\rho_0$ \\
$\gamma$ & 1/3 & Cowan exponent\\
\hline 
\end{tabular} 
\caption{Key parameters used in the X2152 EOS table.}
\label{tab:x2152}
\end{table}

In both X2151 and X2152, the equilibrium conditions were chosen to be in the hexagonal phase, with a density of 2.258 g/cm$^3$, at 295 K and 1 atm. The cold curves are identical in the two models and were fit to calculations from this study and Hugoniot measurements from the Marsh compendium~\cite{MarshLASL1980}. 
Since the ground state phase was taken to be hexagonal the transformation to the cubic phase was represented by employing break-points~\cite{leos2} to transition from the hexagonal cold-curve to the cubic cold-curve at 10 GPa (the wurtzite phase is essentially combined with the cubic phase in this QEOS form).  This transformation pressure is slightly higher than what is reported (1--6~GPa~\cite{BNPD}) but was chosen so that the density where the transformation begins is notably denser than the reference density; this was a practical choice to enhance the stability of the EOS when employed during hydrodynamic simulations.
The first-principles isochores calculated for this work were used to constrain the ion-thermal models; specifically, the density dependent Gr\"uneisen model, and the Cowan liquid model. The largest difference between X2151 and X2152 (outside of the electron-thermal model) is that the best ion-thermal fit for X2151 (hybrid electron-thermal) was found using a Cowan exponent of 0.5, conversely the best fit for X2152 (purely Purgatorio) was determined using the canonical value of 1/3. All other EOS parameters (melt temperature, Debye temperature, etc.) 
were taken directly from known literature.
The thermodynamic parameters in the ion thermal model are determined by fitting the pressure data from PIMC, DFT-MD, and {\footnotesize ACTEX},
taking into account the range of applicability of each method. The key parameters used in X2152 are shown in Table~\ref{tab:x2152}.
In order to avoid problems with energy offsets (energy zeros) in various techniques, only the pressure data are  used for constructing the LEOS tables.
The fidelity of this procedure is discussed here. 

We note that the EOS obtained using different electronic structure theories can vary depending on the underlying physics. For example, orbital-free (OF) MD, which significantly reduces computational cost of standard DFT-MD by constructing the energy functional in a form that is independent of electronic wavefunctions, predicts CH to be less compressible at the compression maximum than predicted by PIMC and Purgatorio~\cite{Zhang2017b,Zhang2018}. Zhang {\it et al.}~\cite{Zhang2018} found that this is because the internal energies calculated by OFMD are lower than PIMC, although the pressures are similar, at the same temperatures. Comparing a recent work~\cite{Danel2018} on carbon EOS using OFWMD (with W standing for Weizs\"acker) to the most recent, Purgatorio-based LEOS 9061 table~\footnote{LEOS 9061 is a multi-phase, Purgatorio-based table for carbon developed by fitting the ionic thermodynamic parameters to the first-principles DFT and PIMC data reported in Ref.~\onlinecite{Benedict_2014}}, the peak compression predicted by OFWMD is also smaller (4.5 by OFWMD versus 4.6 by LEOS 9061). In addition, OFMD calculations for silicon~\cite{Hu2016} shows a single compression maximum along the Hugoniot, whereas PIMC predicts two peaks corresponding to $K$ and $L$ shell ionization respectively.

\begin{figure}
\centering\includegraphics[width=0.23\textwidth]{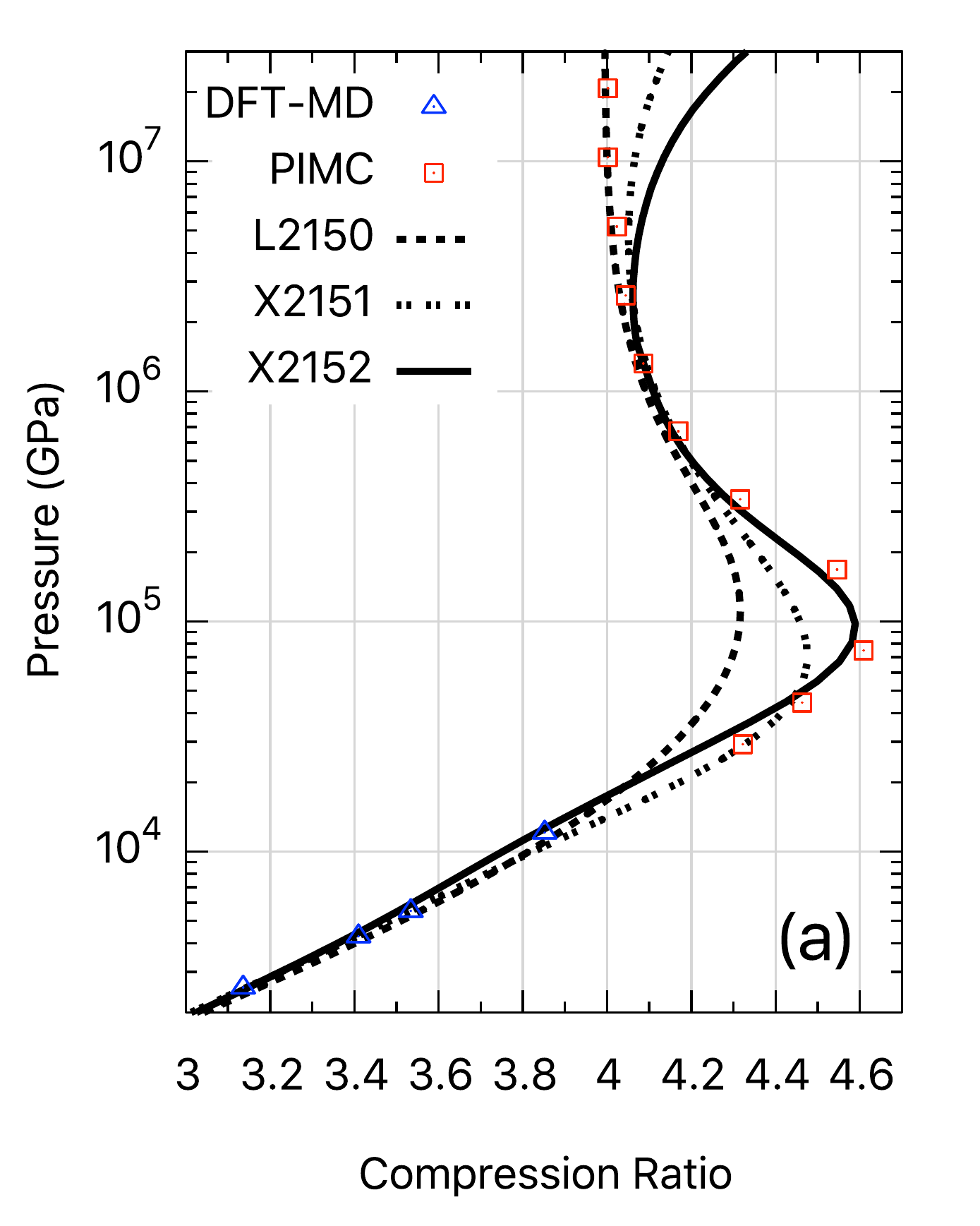}
\centering\includegraphics[width=0.23\textwidth]{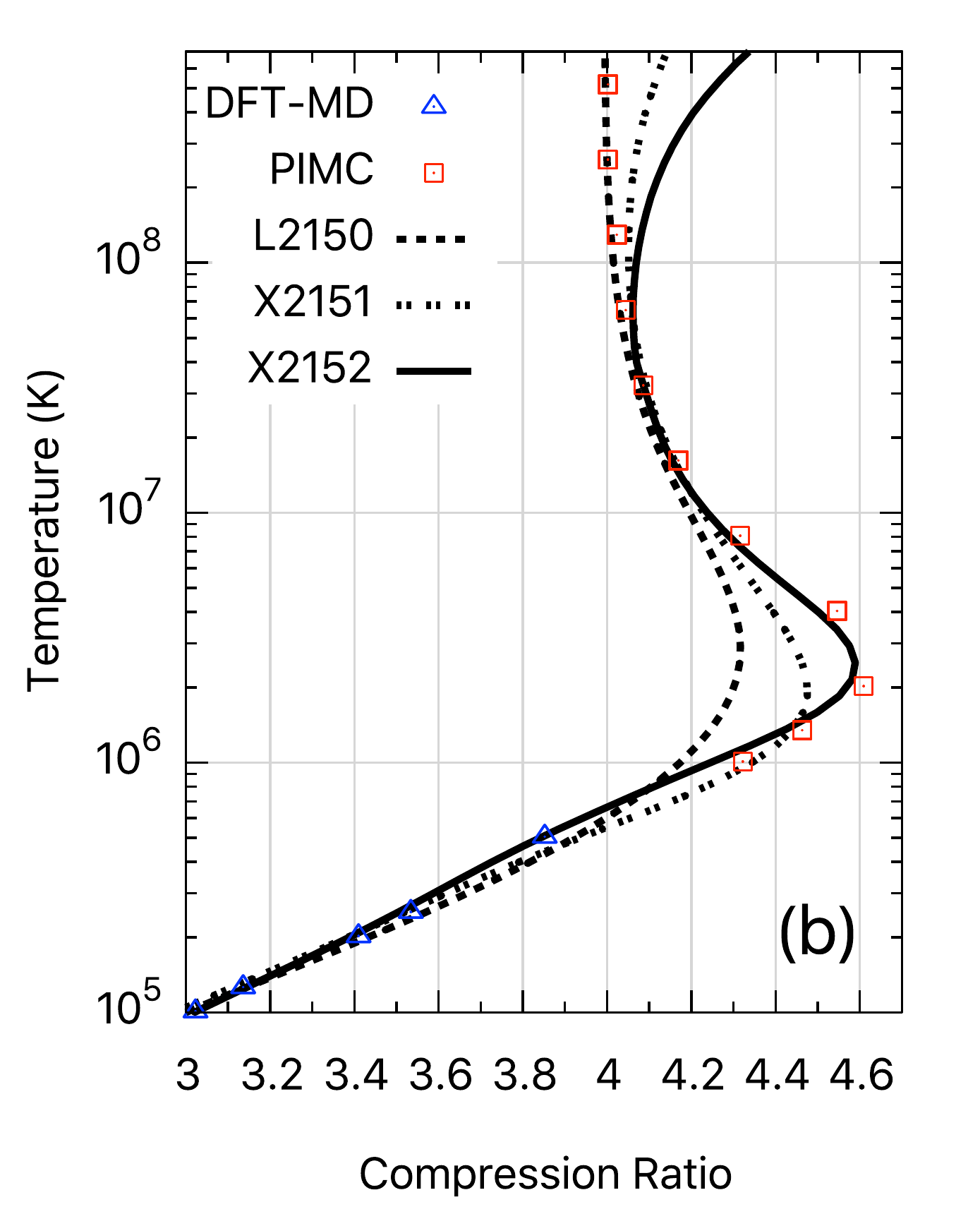}
\caption{\label{fig:bnvschug1} (a) Pressure- and (b) temperature-compression Hugoniot of BN predicted by different LEOS models in comparison with PIMC and DFT-MD (pwPAW). The initial density of all Hugoniot curves are set to be 2.15 g/cm$^3$. Note that the deviations at above $10^6$ GPa and 2$\times10^7$ K are due to the electron relativistic effect, which is included in the Purgatorio tables (thus fully in X2152 and partially in X2151) but not in L2150 or PIMC.}
\end{figure}

We examine the internal energy differences by comparing the Hugoniot curves for BN based on three LEOS tables (LEOS 2150, X2151, and X2152), for which the electron thermal free energy are constructed differently, as we have explained previously in this section.
The results are shown in Fig.~\ref{fig:bnvschug1}. Consistent with previous studies, we find that the TF-based model (L2150) predicts a lower peak compression with a broader shape along the vertical axis than the fully Purgatorio-based model (X2152).  As expected, the model which combines TF and Purgatorio models lies between the two.  Both the shape and the magnitude of the 
peak compression are intimately related to the $K$-shell ionization of B and N.  The TF model is broad due to the neglect of the shell effects, and we observe that the peak compression 
becomes sharper as one accounts for the $K$-shell ionization of B (X2151), and sharper still when we also account for the shell structure of N (X2152).

\begin{figure}
\centering\includegraphics[width=0.45\textwidth]{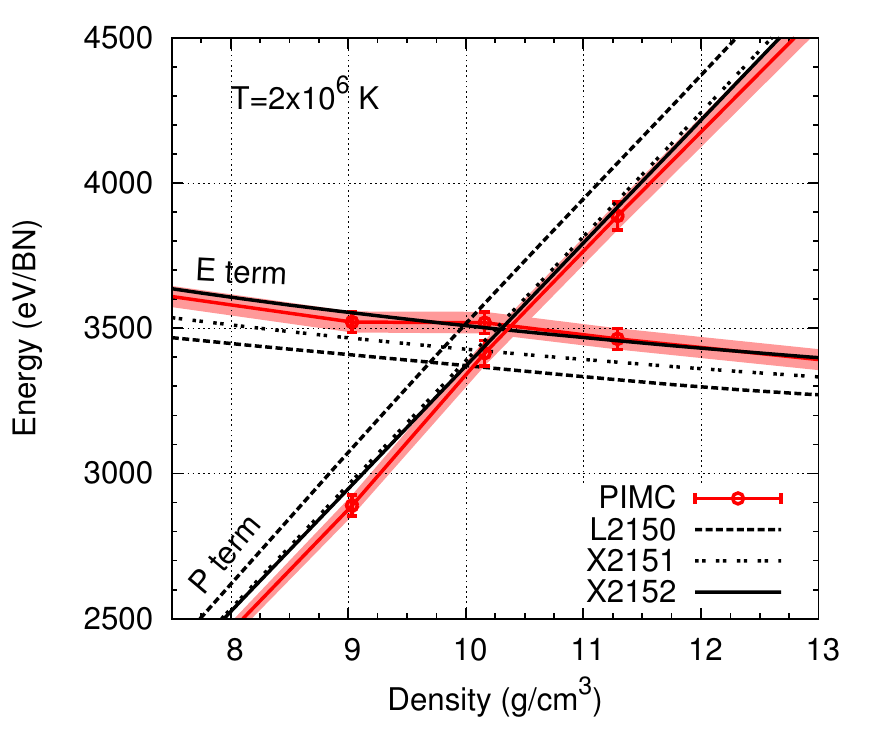}
\caption{\label{fig:bnvschug2} Comparison of the pressure and the energy terms of the Hugoniot function along the 2$\times10^6$ K isotherm, which is near the compression maximum. Shaded areas denote the error bar of the PIMC data.}
\end{figure}

The differences in the maximum compression predicted by the different models can be explained by decomposing the Hugoniot function [left-hand side of the Hugoniot equation $E-E_\text{i}-(P+P_\text{i})(V_\text{i}-V)/2=0$, where $(E,P,V)$ and $(E_\text{i},P_\text{i},V_\text{i})$ denote the energy, pressure, and volume of the sample in the shocked and the initial states, respectively] into the energy term $E-E_\text{i}$ and the pressure term $(P+P_\text{i})(V_\text{i}-V)/2$ and comparing the two as functions of density along isotherms. Figure~\ref{fig:bnvschug2} shows such comparisons at 2$\times10^6$ K, which is near the compression maximum along the shock Hugoniot  (Fig.~\ref{fig:bnvschug1}). The density at which the energy and the pressure curves cross is the Hugoniot density at this temperature. We find that the pressure curves of X2151 and X2152 are on top of each other, but their energies are different. The energies of X2151 are lower, leading to a smaller compression ratio than X2152. In comparison, X2152 data are similar to PIMC in  both energy and pressure. This indicates that when constructing an EOS model by merely fitting pressure, it is important to make the electronic contribution fully Purgatorio-based. This is not surprising because Purgatorio is essentially a DFT method. 
The EOS consistency here demonstrates that the agreement in EOS between PIMC and DFT is not accidental, but represents a consistent description of the electronic interaction in both methods.
In addition, Fig.~\ref{fig:bnvschug2} shows the non-smoothness and error bar of the PIMC data at 2$\times10^6$ K, which leads to an uncertainty in the compression ratio of $\lesssim$0.05 (or $\lesssim$1\%). This represents the level of uncertainty in our reported compression maximum along the Hugoniot by PIMC. At both higher and lower temperatures, the uncertainties are smaller because of the smaller error of the EOS data and higher smoothness of the data along isotherms.

\section{Results and Discussion}\label{sec:results}

\subsection{Isochore Comparisons}\label{subsec:beoscompare}

\begin{figure}
\centering\includegraphics[width=0.48\textwidth]{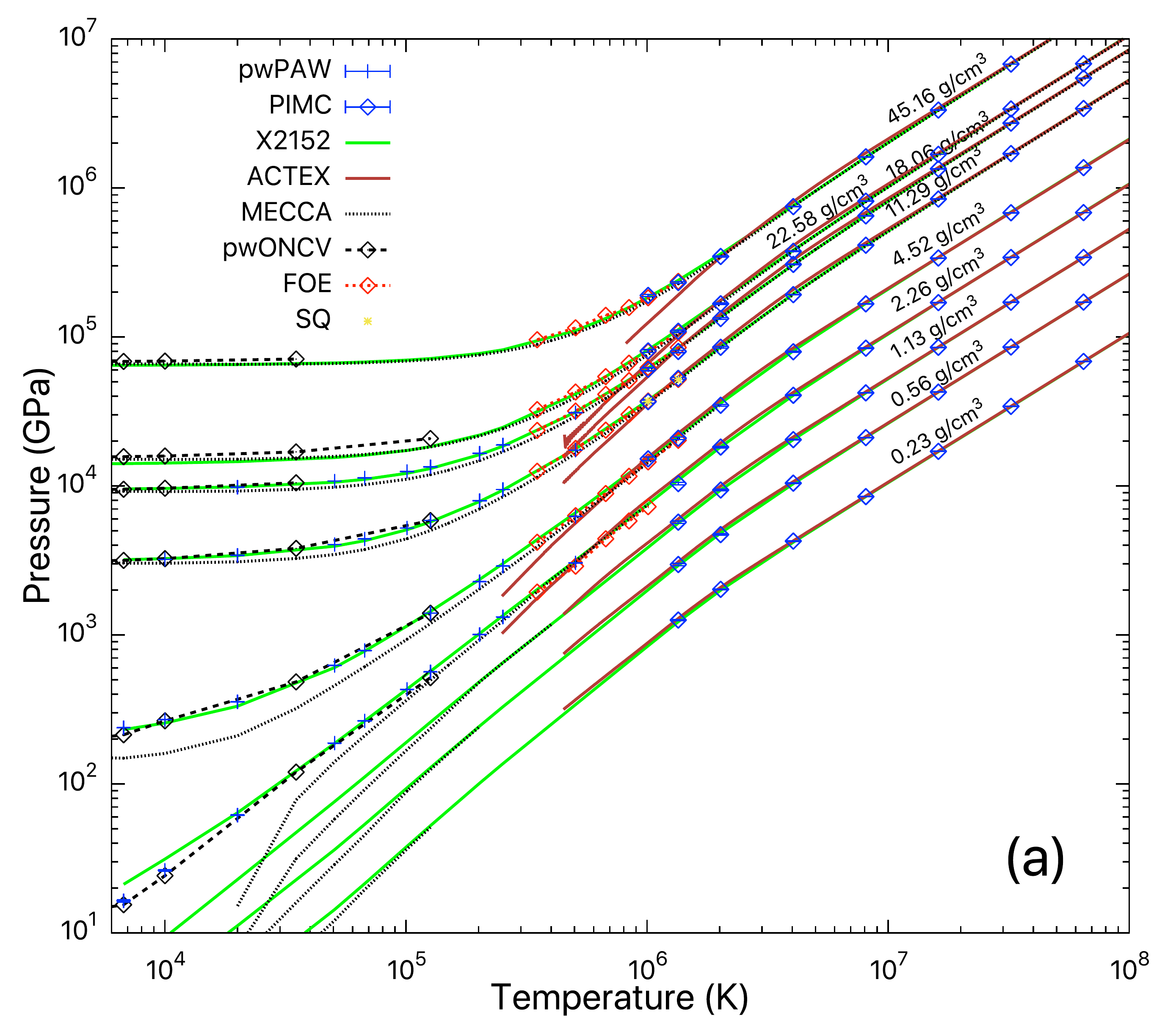}
\centering\includegraphics[width=0.48\textwidth]{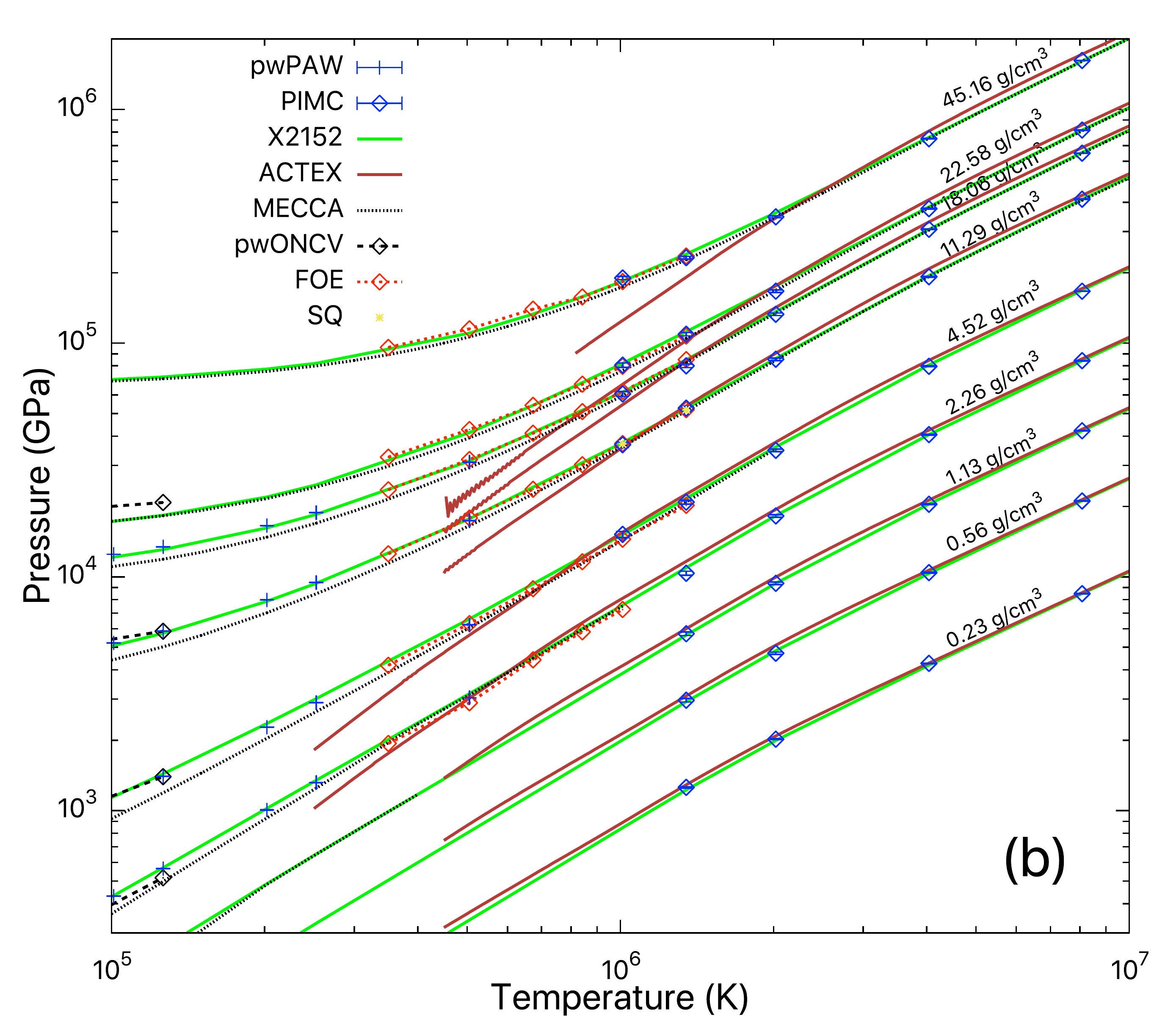}
\caption{\label{fig:pisochore2} Comparison of the pressure-temperature profiles of BN along several isochores from PIMC, DFT-MD (PAW, frozen 1s), DFT-MD (ONCV, frozen 1s), FOE (all-electron), SQ (all-electron), ACTEX, MECCA, and X2152. Subplot (b) is a zoom-in version of (a).}
\end{figure}

In order to evaluate the performance of recent extensions of DFT methods to high temperature, we compare the computed EOS data from 
PIMC, pwPAW, pwONCV, FOE, SQ, {\footnotesize ACTEX}, and {\footnotesize MECCA}. We choose the X2152 model along several 
isochores between 0.23 and 45.16~g/cm$^3$ in Fig.~\ref{fig:pisochore2} for the basis of performing the comparison. 
Figure~\ref{fig:pisochore2}(b) highlights the comparison in the temperature range of 10$^5$--10$^7$~K.
This is the regime where $1s$ electrons are significantly ionized, providing an important testbed for different methods.

\begin{figure}
\centering\includegraphics[width=0.48\textwidth]{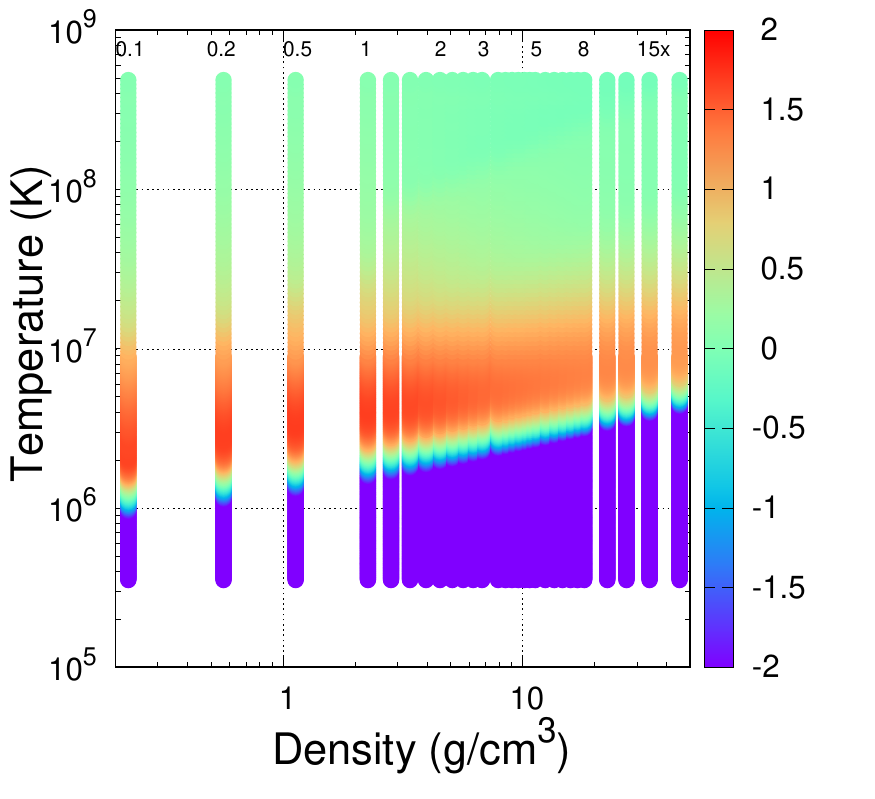}
\caption{\label{fig:eisochore} Percent difference in internal energy of BN between ACTEX and X2152 along several isochores. The compression ratio (with respect to $\rho_0$=2.258 g/cm$^3$) are labeled at the top of the plotting area. The reference points for ACTEX and X2152 are both at $\rho_0$ and ambient temperature.}
\end{figure}

We find that, at temperatures greater than 2$\times10^6$~K, PIMC, {\footnotesize ACTEX}, and {\footnotesize MECCA} results show excellent agreement with each other, while the {\footnotesize ACTEX} predictions are slightly higher than the other two methods only at higher densities. At densities above 4.52 g/cm$^3$ and temperatures below 1.35$\times10^6$ K, deviations of {\footnotesize ACTEX} from the other methods are evident, which indicates a cut-off temperature ($T_\text{cutoff}$) below which the {\footnotesize ACTEX} method breaks down. This is where the two-body term at order 2 in the activity becomes comparable to the Saha term, 
which we use as a simple measure of the point where higher order terms start to contribute. Since those terms are not included in {\footnotesize ACTEX}, we can consider this to be the limit of the current theory.
Moreover, we have plotted the percent differences between {\footnotesize ACTEX} and X2152 data (see Fig.~\ref{fig:eisochore} for the comparison in energy; pressure plots look similar), and found the cutoff is dependent on the density:  $T_\text{cutoff}$ gradually increases from 10$^6$ K to 4$\times10^6$ K as density increases from 0.1- to 20-times $\rho_0$. Above $T_\text{cutoff}$, the agreement between {\footnotesize ACTEX} and X2152 data is excellent, with differences below 2\% in general.

\begin{figure*}
\centering\includegraphics[width=0.85\textwidth]{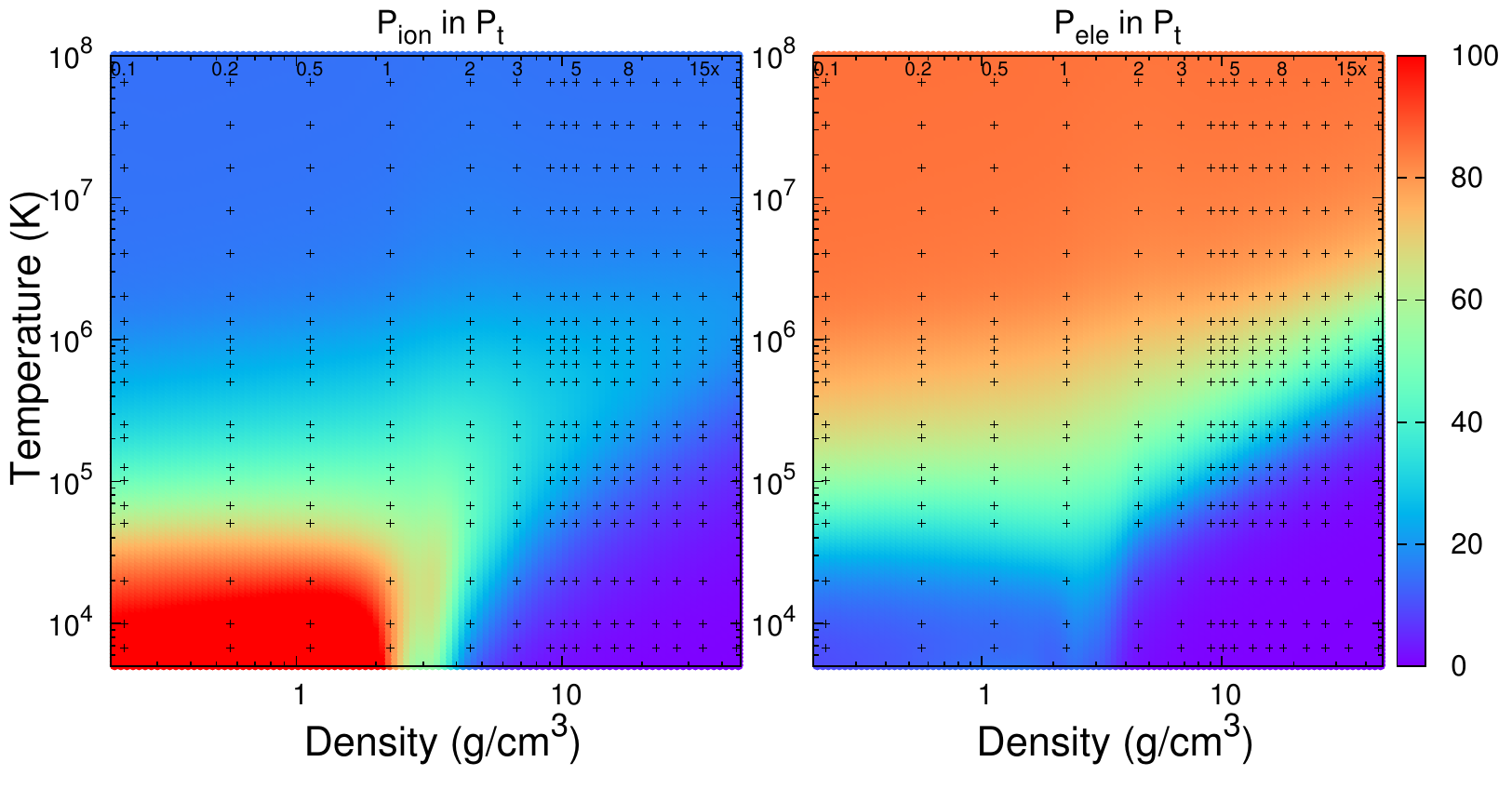}
\caption{\label{fig:bnpcomp} Percent contributions of the ion thermal (left) and electron thermal (right) terms to the total pressure of BN. The remaining contributions are from the cold curve. The temperature-density conditions corresponding to several isochores along which we performed EOS calculations are shown with '+' symbols.}
\end{figure*}

Our pressure-temperature profiles by {\footnotesize MECCA} are overall consistent with those by PIMC, pwPAW, pwONCV, FOE, SQ, and {\footnotesize ACTEX}. The agreement is best at densities higher than 4.5~g/cm$^3$ and temperatures higher than $10^6$~K, where the contributions to the EOS from the ions 
(the ion thermal contributions) are less significant than those from the thermal electrons (see Fig.~\ref{fig:bnpcomp}).  

\begin{figure}
\centering\includegraphics[width=0.5\textwidth]{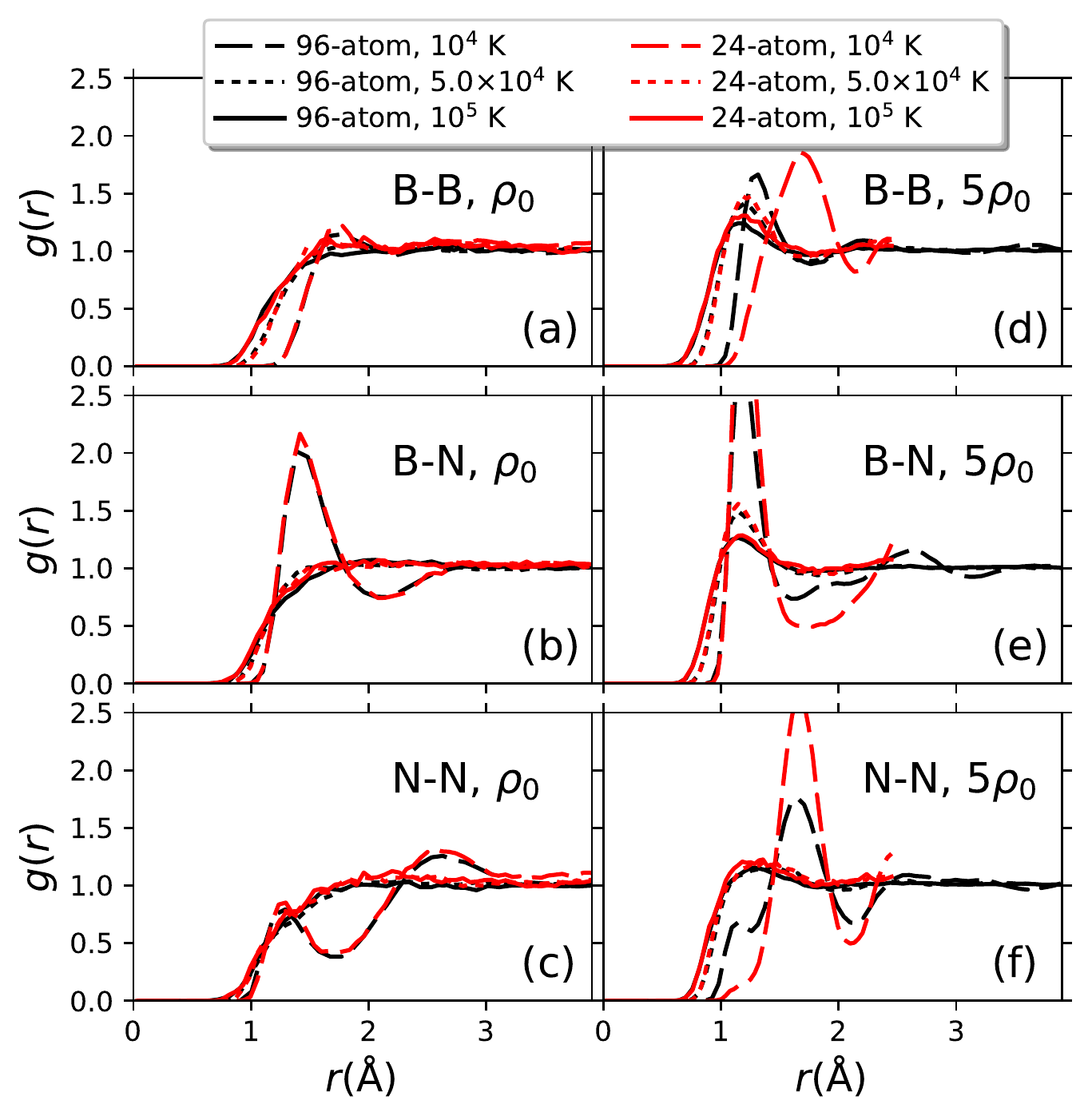}
\caption{\label{fig:gr}  Comparison of the nuclear pair correlation function obtained from DFT-MD (pwPAW) for BN using 24-atom (red) and 96-atom (dark) cells at two different densities and three temperatures. The reference density $\rho_0$ is 2.26 g/cm$^3$. The peaks at $10^4$ K indicates a polymeric structure of the liquid. Differences between small and large cells are evident at 4000 K, indicating a significant finite size effect. This effect is stronger at higher densities and becomes negligible at temperatures higher than 5$\times10^4$ K.}
\end{figure}

At intermediate-low densities (0.23-2.3~g/cm$^3$), we observe a discrepancy between {\footnotesize MECCA} and the DFT-MD/X2152 data, and it grows larger as temperature decreases further below $10^5$ K. This is because the {\footnotesize MECCA} simulations are performed using static configurations with 2 atoms in the B2 (cesium chloride) structure, which do not include ion motion, and we have thus approximated the ion thermal effect by adding ideal gas corrections to the pressures and energies. However, at the low-temperature conditions, the nuclei show significant correlations by forming polymers, such as N-N pairs or B-N structures that are characterized by the strong fluctuations in the radial pair distribution function at $10^4$~K and shown in Fig.~\ref{fig:gr}(a)-(c). Therefore, by disregarding the vibrational and rotational contributions, the ideal gas model underestimates the EOS at these conditions. 
As temperature exceeds $5\times10^4$~K, the features in the pair distribution function quickly smooth out because the polymeric structures are de-stabilized by thermal effects, which makes the ideal gas approximation for the ions work better and explains the improved agreement between the EOS from DFT-MD and {\footnotesize MECCA}.
Moreover, we note that the agreement between the EOS from X2152 and {\footnotesize MECCA} can be improved by replacing the ideal-gas correction with the ion thermal model from X2152. The differences at $\rho>\rho_0$ reduce more by applying a constant shift to the {\footnotesize MECCA} pressures to anchor the pressure-zero point at $\rho_0$ and 300 K. These findings explain the good consistency between the shock Hugoniot predicted by X2152 and {\footnotesize MECCA} EOS data, which we address in Sec.~\ref{subsetheorycompare}.

At densities higher than 2.26 g/cm$^3$, the radial distribution function also show significant pair correlations at temperatures below $10^5$ K (Fig.~\ref{fig:gr}(d)-(f)). However, the agreement between the EOS from {\footnotesize MECCA} and those from DFT-MD are far better than at lower densities. This is the regime where the cold curve contribution dominates the EOS, as Fig.~\ref{fig:bnpcomp} implies.
The excellent agreement between {\footnotesize MECCA} and DFT-MD EOS indicates the effects of the simulation cell and the non-ideal ion thermal contribution are less significant in the more strongly compressed ($\rho\geq5\times\rho_0$) regime.

At 2.26 g/cm$^3$ and $T<2\times10^4$ K, We also observe differences between X2152 and DFT-MD. 
This can be explained by the differences in the cold curve between X2152 and DFT-MD. The energy minimum in X2152 is set to $\rho_0=2.26$ g/cm$^3$ corresponding to h-BN, while DFT-MD tends to stabilize c-BN because of the cubic simulation cell being implemented for the liquid simulations. 
In fact, we found that altering the cold-curve in X2152 such that the $\rho_0$ is more in line with the ambient density of c-BN allows for better agreement with these low temperature points.

\begin{figure}
\centering\includegraphics[width=0.45\textwidth]{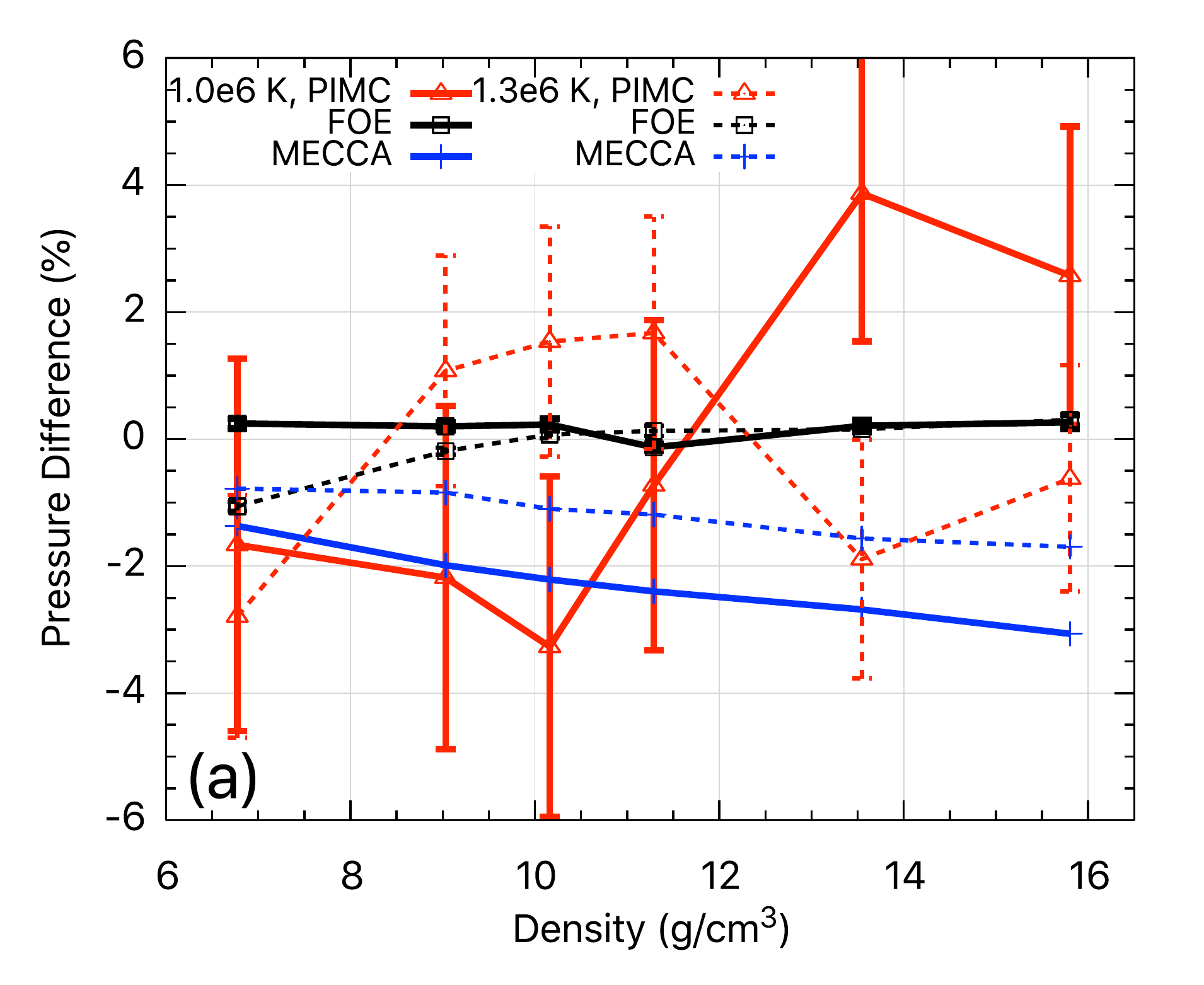}
\centering\includegraphics[width=0.45\textwidth]{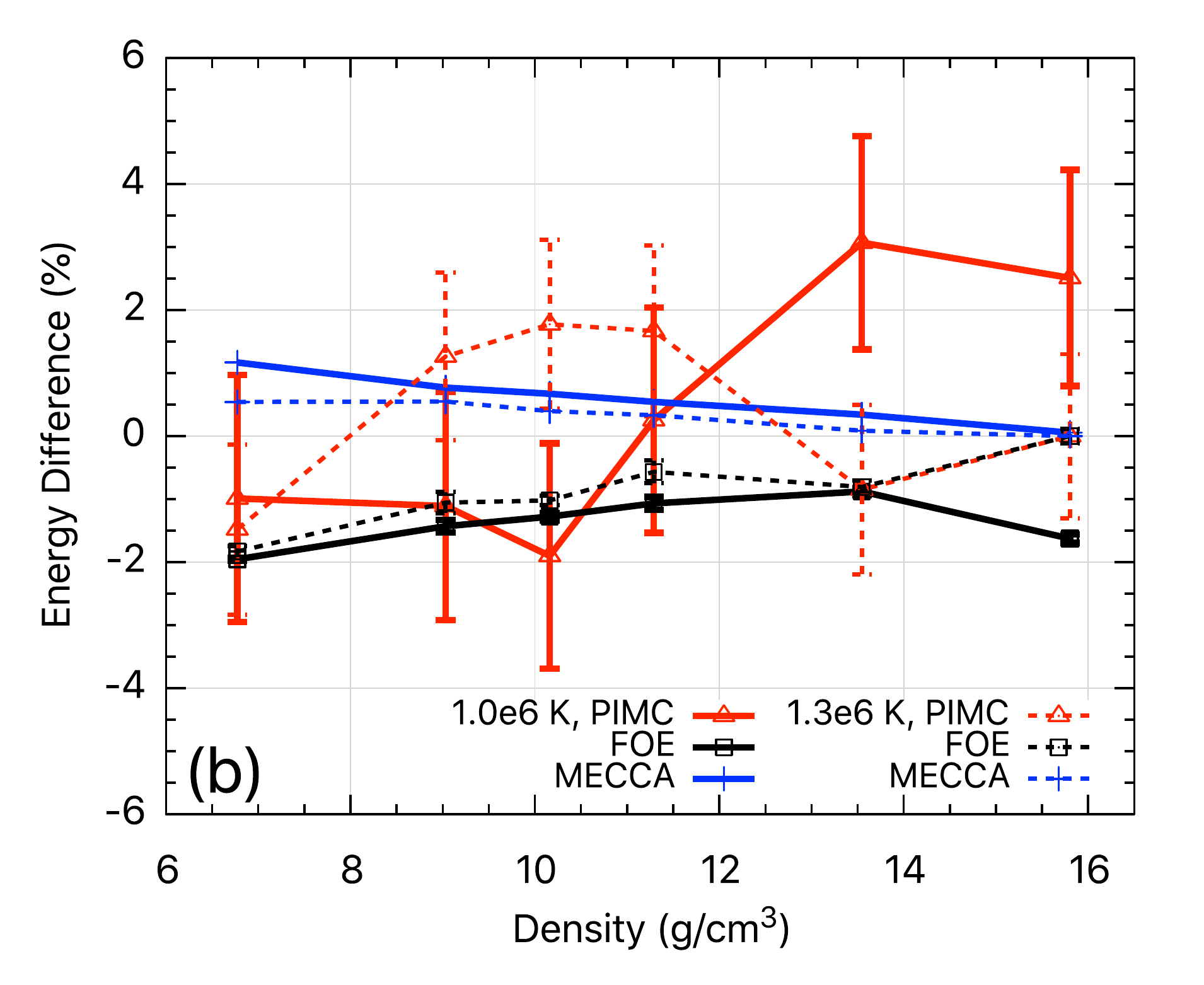}
\caption{\label{fig:sqdftvsfoevspimc} EOS differences of PIMC (red), FOE (black), and MECCA (blue) relative to SQ along two isotherms (1.01$\times10^6$ and 1.35$\times10^6$ K). Because of the different references chosen in the EOS datasets, all energies have been shifted by the corresponding value at 15.80~g/cm$^3$ and 1.35$\times10^6$~K. The energy differences are normalized by the corresponding ideal gas values ($21k_BT$ per BN).
The statistical error bars correspond to the 1$\sigma$ uncertainty of the FOE and PIMC data.}
\end{figure}

\begin{table*}
\scriptsize
\centering
\begin{tabular}{cc|cc|cc|cc|cc}
\hline 
 &  &  \multicolumn{2}{c|}{SQ} &  \multicolumn{2}{c|}{FOE}   & \multicolumn{2}{c|}{PIMC} & \multicolumn{2}{c}{MECCA}\\
 \hline
$\rho$ & $T$ & $P$ & $E$ & $P$ & $E$ & $P$ & $E$ & $P$ & $E$ \\ 
 (g/cm$^3$) & (K) & (GPa) & (Ha/BN) & (GPa) & (Ha/BN) & (GPa) & (Ha/BN) & (GPa) & (Ha/BN) \\
\hline 
6.77 & 1010479 & 21807$\pm$9 & -18.375$\pm$0.016 & 21860$\pm$15 & -19.688$\pm$0.019 & 21446$\pm$628 & -19.039$\pm$1.317 & 21510 & -17.589 \\
9.03 & 1010479 & 29297$\pm$12 & -20.045$\pm$0.018 & 29355$\pm$18 & -21.006$\pm$0.069 & 28664$\pm$775 & -20.791$\pm$1.217 & 28721 & -19.528 \\
10.16 & 1010479 & 33136$\pm$17 & -20.664$\pm$0.021 & 33212$\pm$29 & -21.522$\pm$0.040 & 32070$\pm$860 & -21.941$\pm$1.201 & 32411 & -20.212 \\
11.29 & 1010479 & 37027$\pm$15 & -21.149$\pm$0.017 & 36979$\pm$20 & -21.866$\pm$0.069 & 36758$\pm$956 & -20.978$\pm$1.201 & 36149 & -20.783 \\
13.55 & 1010479 & 44946$\pm$23 & -21.921$\pm$0.022 & 45040$\pm$40 & -22.511$\pm$0.040 & 46718$\pm$1087 & -19.857$\pm$1.138 & 43755 & -21.692 \\
15.80 & 1010479 & 53176$\pm$39 & -22.379$\pm$0.032 & 53317$\pm$58 & -23.472$\pm$0.046 & 54562$\pm$1281 & -20.691$\pm$1.152 & 51570 & -22.342 \\
6.77 & 1347305 & 31097$\pm$12 & 7.553$\pm$0.020 & 30769$\pm$20 & 5.913$\pm$0.079 & 30240$\pm$577 & 6.226$\pm$1.210 & 30855 & 8.040 \\
9.03 & 1347305 & 41369$\pm$15 & 4.580$\pm$0.019 & 41291$\pm$22 & 3.634$\pm$0.150 & 41816$\pm$759 & 5.713$\pm$1.190 & 41022 & 5.073 \\
10.16 & 1347305 & 46621$\pm$18 & 3.528$\pm$0.022 & 46654$\pm$27 & 2.613$\pm$0.066 & 47342$\pm$858 & 5.119$\pm$1.201 & 46111 & 3.884 \\
11.29 & 1347305 & 51838$\pm$26 & 2.565$\pm$0.029 & 51904$\pm$41 & 2.057$\pm$0.160 & 52711$\pm$964 & 4.061$\pm$1.212 & 51226 & 2.863 \\
13.55 & 1347305 & 62537$\pm$22 & 1.137$\pm$0.021 & 62633$\pm$42 & 0.415$\pm$0.090 & 61365$\pm$1153 & 0.378$\pm$1.206 & 61566 & 1.215 \\
15.80 & 1347305 & 73360$\pm$30 & 0.000$\pm$0.024 & 73582$\pm$59 & 0.000$\pm$0.101 & 72905$\pm$1299 & 0.000$\pm$1.166 & 72125 & 0.000 \\
\hline 
\end{tabular} 
\caption{Comparison of computed internal energies and pressures from SQ, FOE, PIMC, and MECCA. The energies have been shifted by setting the reference to their respective values at 15.80 g/cm$^3$ and 1.35$\times10^6$ K, at which the pressures are close to each other. The errors in the SQ, FOE, and PIMC data are the statistical 1$\sigma$ error bar determined by blocking analysis~\cite{Allen1987}.}
\label{tab:foevssqvspimc}
\end{table*}

We compare the EOS data from SQ with those from PIMC, FOE, and {\footnotesize MECCA} along two different isotherms: 1.01$\times10^6$ and 1.35$\times10^6$ K. Their values are listed in Tab.~\ref{tab:foevssqvspimc} and the differences shown in Fig.~\ref{fig:sqdftvsfoevspimc}(a) and (b) for pressures and energies, respectively.
Our FOE and SQ pressures are in excellent agreement with each other (differences are less than 1\%). This can be explained by the use of all-electron ONCV potentials and the DFT-MD nature of both methods. The FOE energies are slightly lower than the SQ values by 1-2\% of the corresponding ideal gas values. 
The small differences can be attributed mainly to different discretization errors in the two approaches, whereas differences associated with trajectory lengths, pseudopotentials, and exchange-correlation functionals were determined to be an order of magnitude smaller.

Our PIMC data at these temperatures scatter around the DFT values, because of the longer paths and larger error bars 
at such conditions. The differences between PIMC and SQ are $<$4\% in pressure and $\lesssim1$ Ha/atom (or $\lesssim3\%$ when normalized by the ideal gas value) in energy, which is typical of what we found about differences between PIMC and DFT-MD in previous work on B~\cite{Zhang2018b} and hydrocarbon systems~\cite{Zhang2018,Zhang2018b}.
{\footnotesize MECCA} data also agree with SQ and FOE at these conditions, with differences $<3\%$ in pressure and $<0.4$ Ha/atom (or $<1.5\%$ when normalized by corresponding ideal gas values) in energy. The cross validation of the different DFT methods and their consistency with PIMC predictions strongly suggest both the PIMC and the DFT-MD approaches, albeit carrying approximations in each, are reliable for studying the EOS of warm dense matter.

Figure~\ref{fig:sqdftvsfoevspimc} and Table~\ref{tab:foevssqvspimc} also show the standard error bars of our EOS data, determined by statistical averaging of the MD (for FOE and SQ) or PIMC data blocks. At the temperatures of 1.01$\times10^6$--1.35$\times10^6$~K, PIMC errors are 2--3\% in pressure and $\sim$0.6 Ha/atom in energy; FOE errors are 0.05--0.8\% in pressure and 0.01--0.08~Ha/atom in energy. In comparison, the statistical error bars of the SQ data are significantly smaller (see Tab.~\ref{tab:foevssqvspimc}). These results, for the first time, establish SQ as an accurate method capable of calculating the EOS of partially ionized, warm-dense plasmas with high precision and accuracy comparable to PIMC.

\subsection{Comparison between theory and experiment}\label{subsec:theoryvsexpt}

\begin{figure*}
\centering\includegraphics[width=0.85\textwidth]{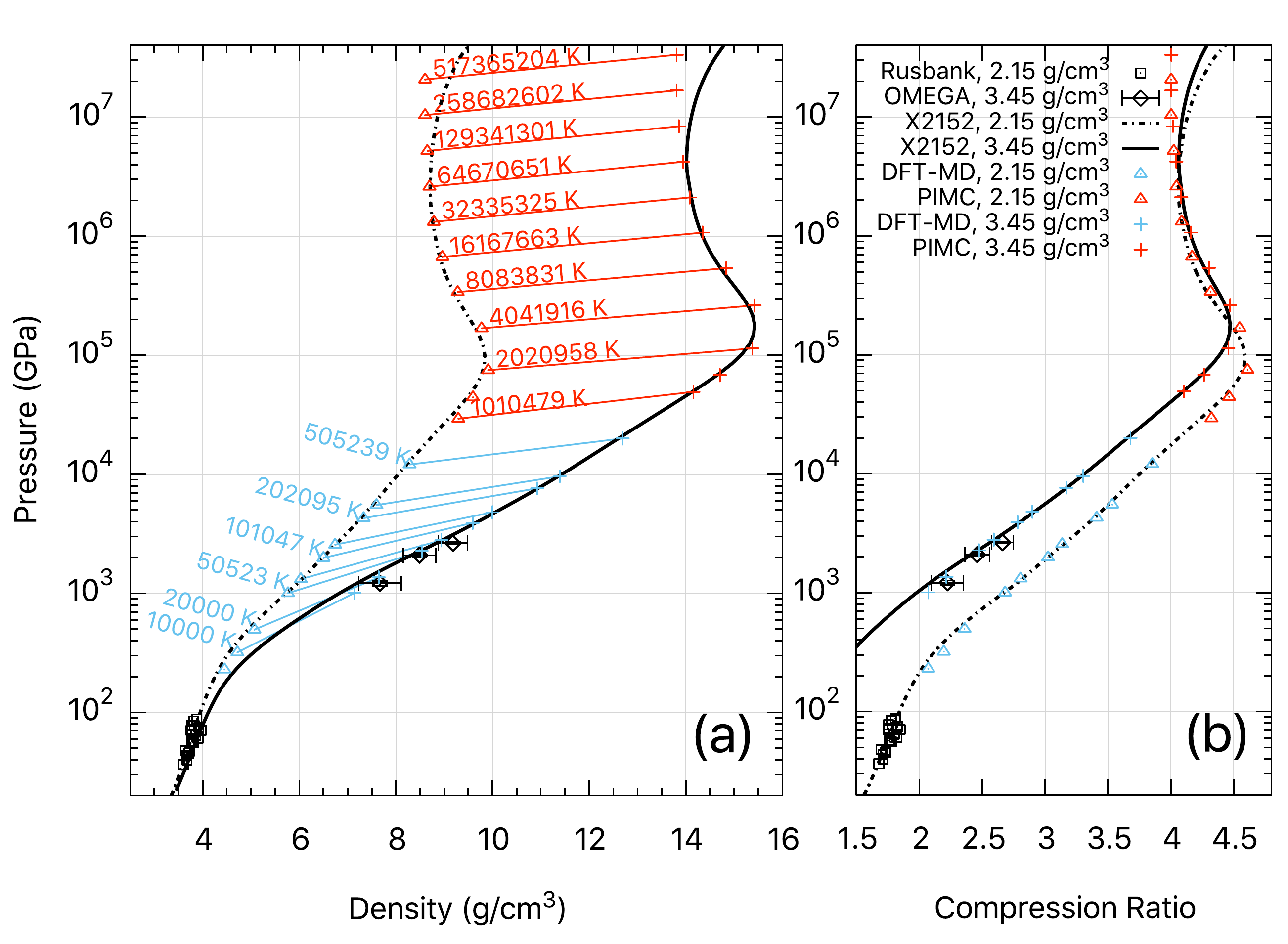}
\caption{\label{fig:prhohugexpt} Comparison of the Hugoniot  of BN from experiment to predictions from PIMC and DFT-MD (pwPAW) simulations and the X2152 model in (a) pressure-density and (b) pressure-compression ratio representations. The initial densities of corresponding Hugoniots are shown in the legend. In (a), equal-temperature conditions along the two Hugoniot curves are connected with lines (as guides to the eyes) to approximate the location of isotherms. The corresponding temperatures are labeled in colored texts. Note that the deviation between PIMC and X2152 curves at above $10^6$ GPa is due to the electron relativistic effect, which is considered in X2152 but not in PIMC.}
\end{figure*}


In this section, we compare our experimental measurements of the pressure-density relation of BN with our theoretical predictions. The experimental data are along the Hugoniot curve, which varies depending on the properties of the sample material. Figure~\ref{fig:prhohugexpt} compiles the experimental and theoretical Hugoniot curves corresponding to two different initial densities ($\rho_\text{i}$): Omega data with $\rho_\text{i}$ of 3.45 g/cm$^3$ and the Rusbank data~\cite{MarshLASL1980} with $\rho_\text{i}$ of 2.15~g/cm$^3$. The corresponding theoretical predictions by X2152 are shown with dark curves. We also show the PIMC and the DFT-MD predictions for 3.45~g/cm$^3$ and 2.15~g/cm$^3$.

The comparison in Fig.~\ref{fig:prhohugexpt} shows very good consistency between the measurements and the theoretical predictions. 
Assisted by the theoretical predictions, we are able to estimate Hugoniot temperatures for the experimental data. We label the Hugoniot temperatures for selected DFT-MD data points with blue-colored text in Fig.~\ref{fig:prhohugexpt}. We find the Omega data points are in the temperature range of 10$^4$--10$^5$ K. 
Our results also show that the PIMC and DFT-MD predicted Hugoniot are in remarkable agreement with X2152 for both initial densities, which spans the Hugoniot curves over a wide range in the phase space. This further shows the validity of the fitting and construction procedure and the quality of our X2152 table. Our calculations and the X2152 model predicts BN to have a maximum compression ratio of 4.59 at 9.8$\times10^4$~GPa for $\rho_\text{i}=2.15$~g/cm$^3$ and 4.47 at 1.8$\times10^5$~GPa for $\rho_\text{i}=3.45$~g/cm$^3$.
We also note that the pressure-density Hugoniots predicted by our different tabular models are very similar (see Fig.~\ref{fig:bnvschug1}) at the pressure regime ($10^3$--3$\times10^3$~GPa) explored in our current experiments. We expect future, accurate experiments at higher pressures (e.g., near the compression maximum) to further check our predictions.

\subsection{Comparison of different EOS methods}\label{subsetheorycompare}

\begin{figure}
\centering\includegraphics[width=0.23\textwidth]{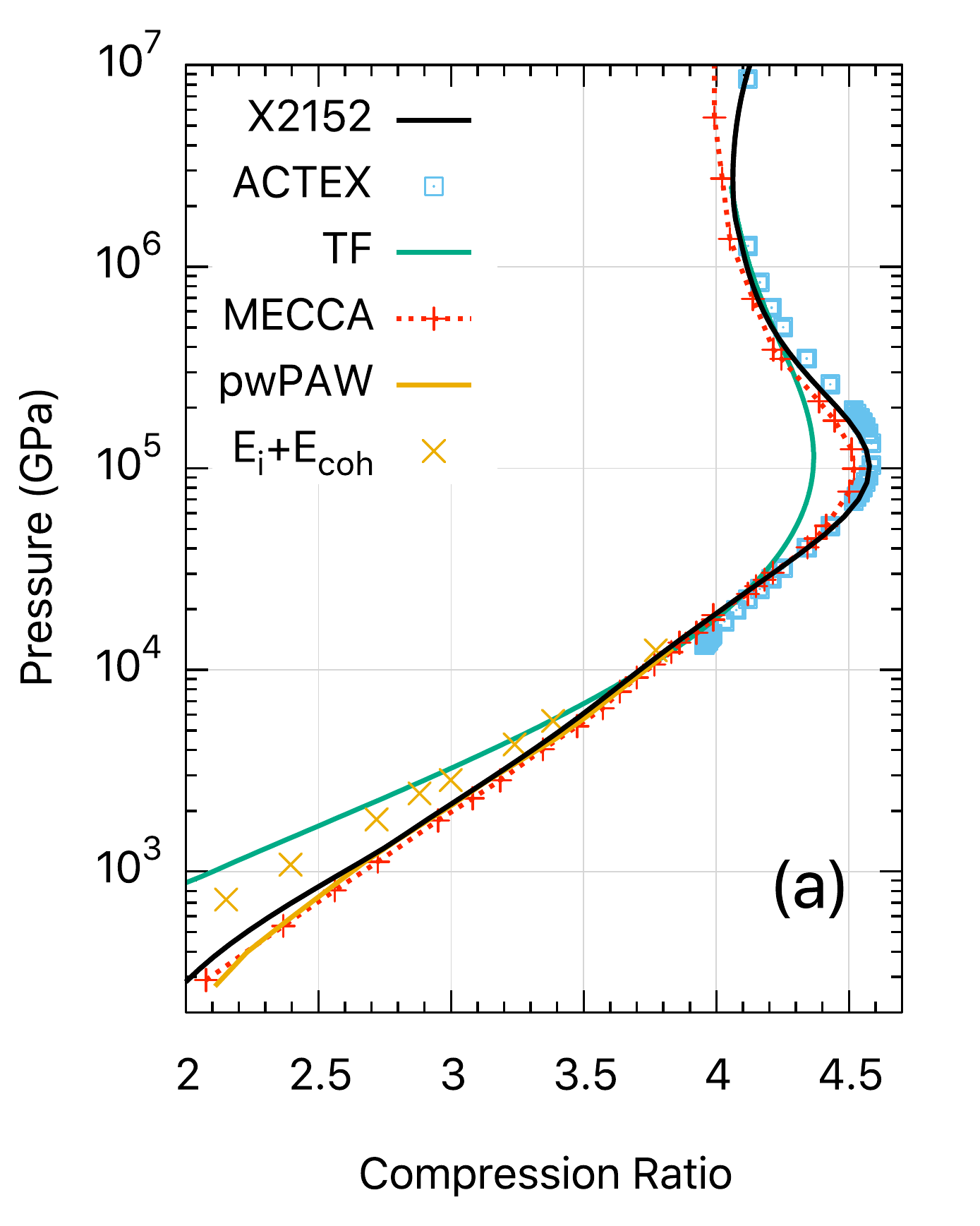}
\centering\includegraphics[width=0.23\textwidth]{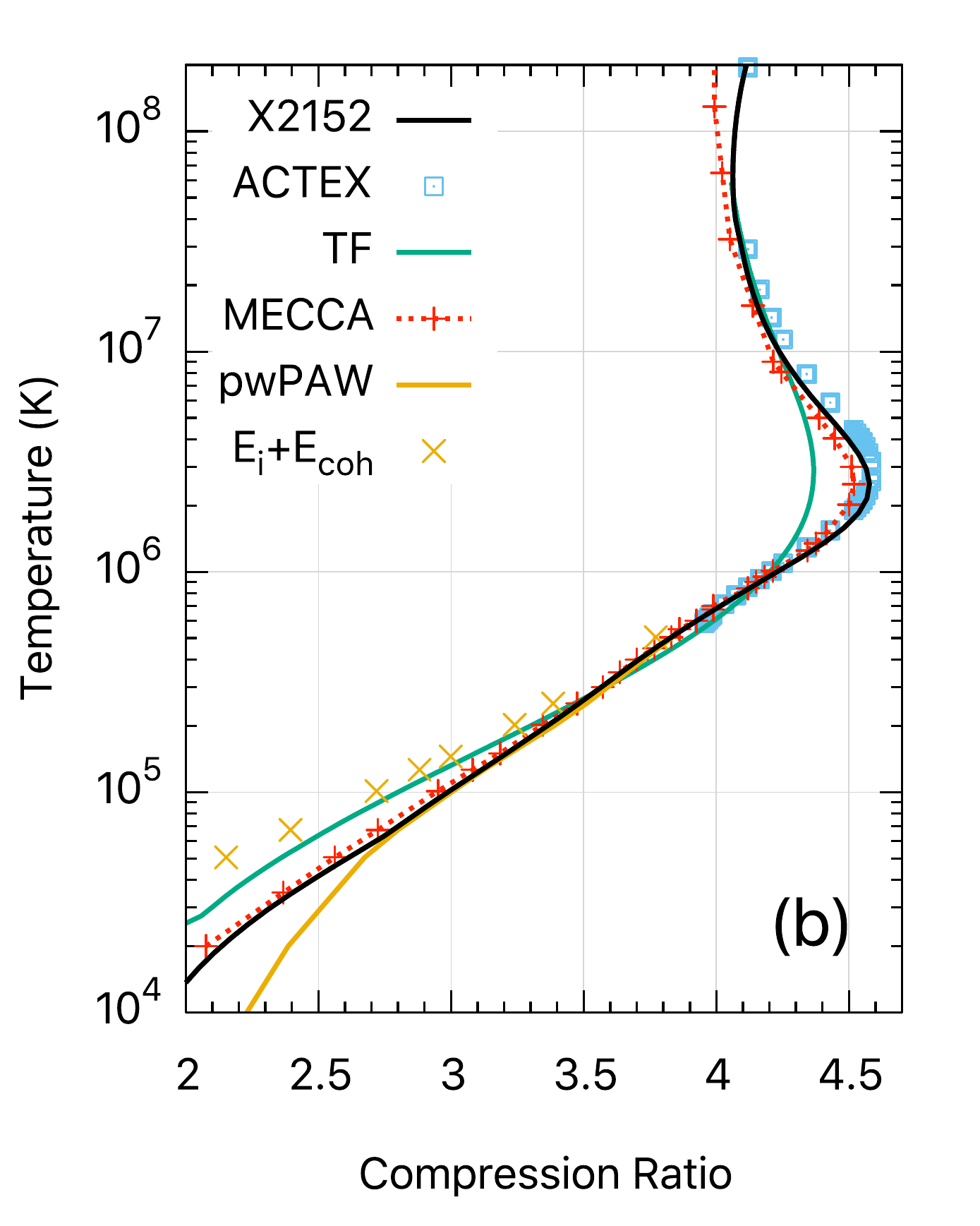}
\caption{\label{fig:prhohugtheory} Comparison of the pressure-compression Hugoniot of BN from different theories and LEOS models. The initial density of every Hugoniot curve is 2.26 g/cm$^3$. Two sets of DFT-MD (pwPAW) Hugoniots constructed with a difference of the cohesive energy ($E_\text{coh}\sim$7.1 eV/atom~\cite{Ooi_2006}) in the initial energy are also shown for comparison.
Note that all MECCA pressures in the EOS have been shifted relative to the value at the initial density and 300 K. Also note that the deviation at above $10^6$ GPa and 2$\times10^7$ K is due to the electron relativistic effect, which is considered in X2152 and ACTEX but not in MECCA.}
\end{figure}

Finally, we make a comprehensive comparison of the shock Hugoniot curves for BN predicted by our different EOS methods. The pressure-compression and temperature-compression Hugoniot curves from {\footnotesize ACTEX}, TF, {\footnotesize MECCA}, and X2152 are shown in Fig.~\ref{fig:prhohugtheory}.  We note that ACTEX and X2152 each intrinsically
accounts for electron relativistic effects, thus the Hugoniot deviates from the nonrelativistic ideal electron gas limit of 4 at very high temperatures ($>10^8$ K).  In comparison, the relativistic correction has not been applied to the TF or MECCA calculations.

At pressures of $\sim10^4$--$10^6$ GPa and temperatures $\sim3\times10^5$--$2\times10^7$ K, {\footnotesize ACTEX}, X2152, and {\footnotesize MECCA} yield very similar Hugoniot profiles and a maximum compression of $\sim$4.55 for $\rho_\text{i}$ of 2.26 g/cm$^3$, while the peak is more broadened according to the TF model and the maximum compression ratio is lower by $\sim$0.2. The peak is associated with the $K$ shell ionization of B and N, which is smoothed out in the TF model because electronic shell effects are missing in this approach but captured by the other methods. The slightly larger compression predicted by {\footnotesize ACTEX} than X2152 is consistent with the $\lesssim2\%$ larger values of the {\footnotesize ACTEX} EOS data than X2152 (Figs.~\ref{fig:pisochore2} and ~\ref{fig:eisochore}). The slightly lower compression predicted by {\footnotesize MECCA} than X2152 can be explained by the non-perfect reconciliation in pressure and energy terms in the Hugoniot function ({\footnotesize MECCA} pressures are slightly lower while energies are similar in comparison to SQ and PIMC, as shown in Fig.~\ref{fig:sqdftvsfoevspimc}). 

In the low-temperature condensed matter regime, we find that, with a constant pressure shift in the EOS, our {\footnotesize MECCA} predictions for the Hugoniot are in good consistency with those of X2152. This indicates the efficacy of using the ideal gas model to approximate the ion thermal effect when constructing EOS using small-size, fixed-lattice models (as in {\footnotesize MECCA}). 
Our TF results predict BN to be stiffer in this regime because the initial energy in TF is estimated using an average-atom method (described in Appendix B),
which may be higher than the actual value because of the excess energy release due to bonding.
We also show differences between X2152 and our DFT-MD (pwPAW) predictions, in particular in Hugoniot temperatures (Fig.~\ref{fig:prhohugtheory}(b)). This is because of the EOS differences between h-BN and c-BN that we have elaborated previously in Sec.~\ref{subsec:beoscompare}.

\subsection{EOS and Hugoniot of isoelectronic materials}\label{subsec:isoebnvsc}

 \begin{figure}
\centering\includegraphics[width=0.23\textwidth]{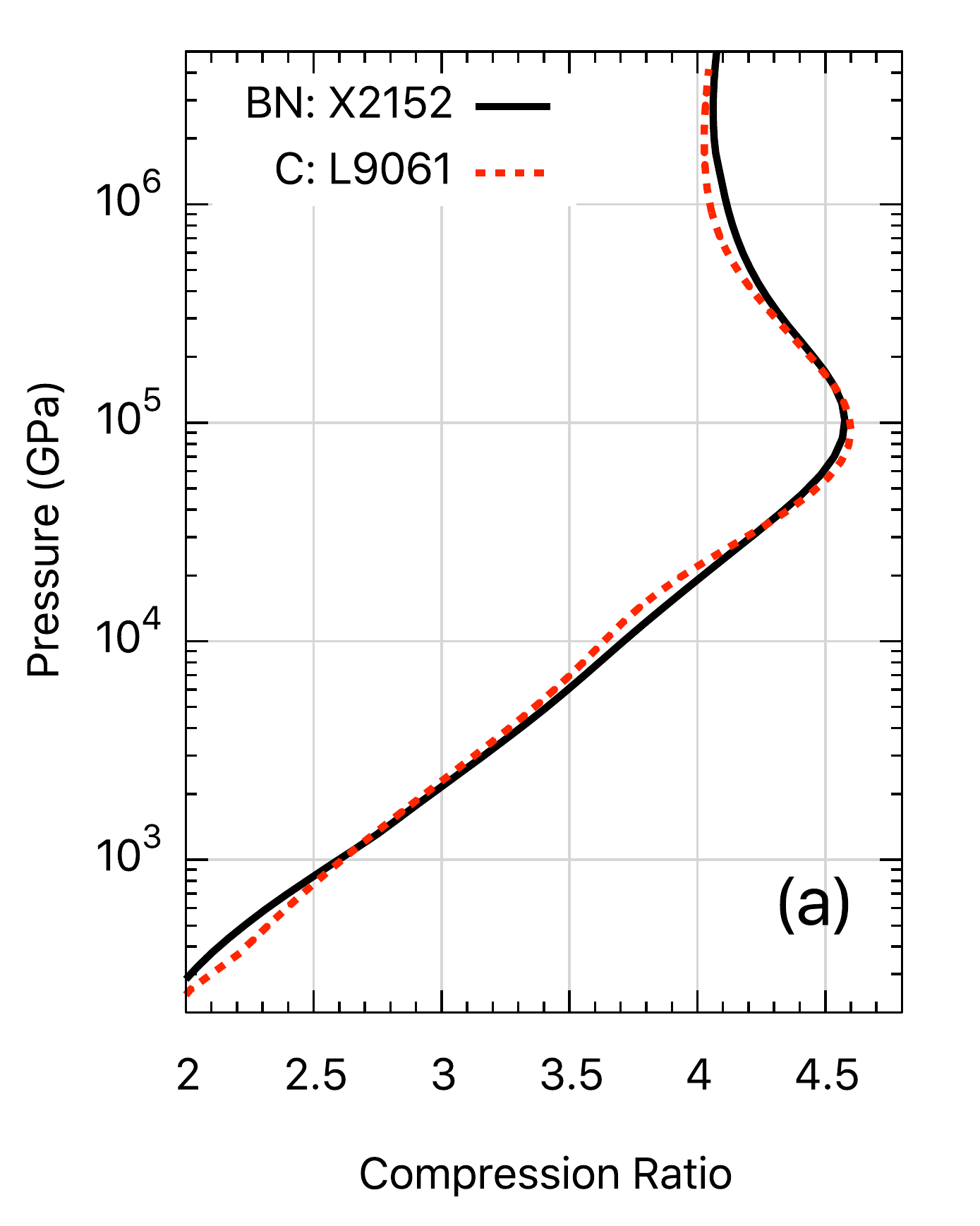}
\centering\includegraphics[width=0.23\textwidth]{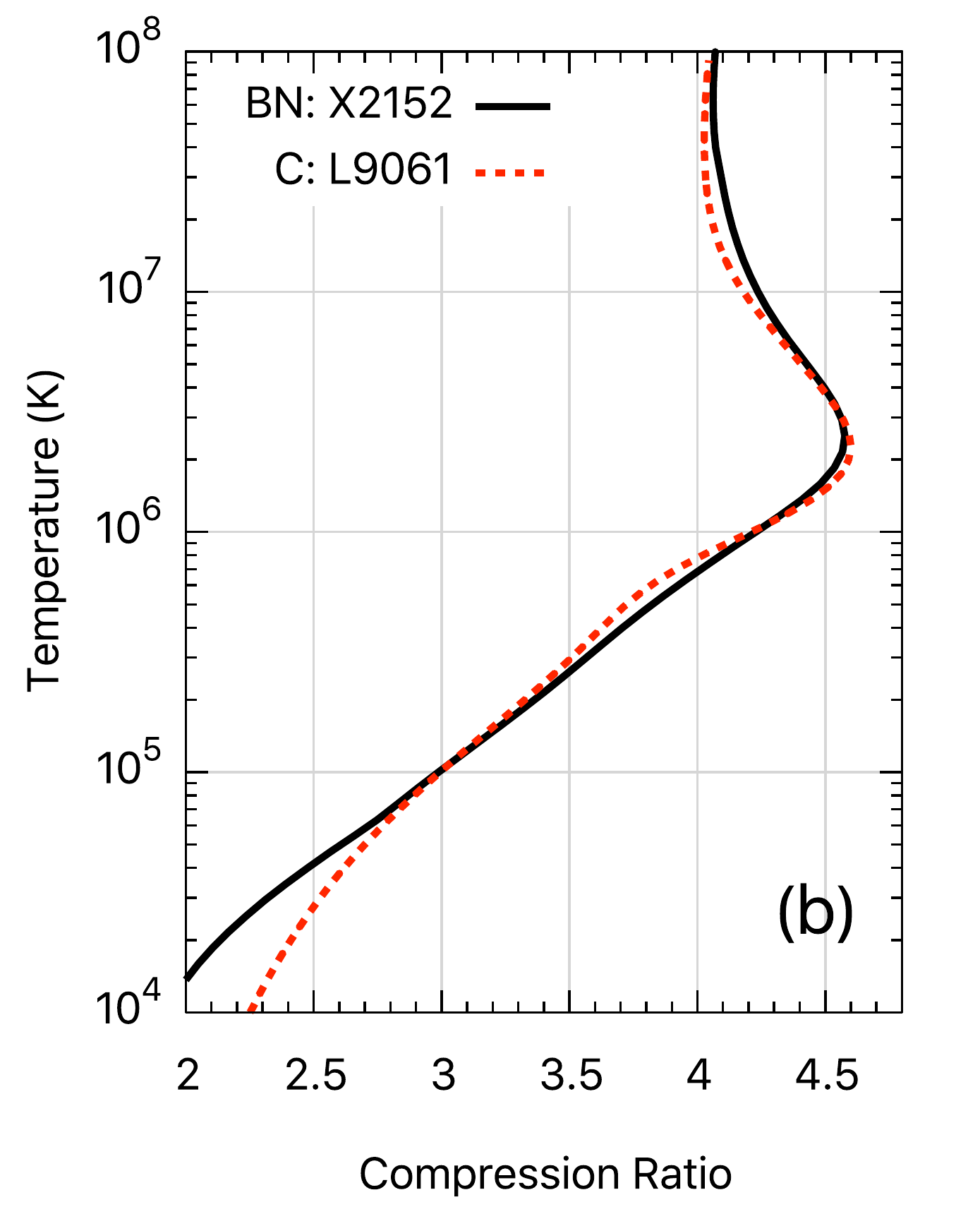}
\caption{\label{fig:bnvschug} (a) Pressure- and (b) temperature-density Hugoniot of BN in comparison with C. The electron thermal contribution to both tables are based on Purgatorio. The initial density of both materials are set to be 2.26 g/cm$^3$.}
\end{figure}

Our EOS models and results for BN enable us to investigate the difference with C---an isoelectronic material of BN. Figure~\ref{fig:bnvschug} compares the Hugoniot of BN and of C based on X2152 and LEOS 9061, setting their initial densities to be the same (2.26~g/cm$^3$). LEOS 9061 is the a multi-phase EOS table constructed for C by using a Purgatorio table for the electron thermal term and fitting DFT and PIMC data~\cite{Benedict_2014} to obtain the ion thermal term, similar to our present work on BN.

The Hugoniot comparison shows that, at temperature regimes of both $10^5$--$10^6$~K and $>10^7$ K, the compression ratio of BN is higher than C. The compression peak is thus slightly narrower for C. This is because the $K$ level of C is in between those of B and N. The differences between BN and C in the low-pressure condensed-matter region ($T<10^5$ K) reflect differences in the cold-curve and ion thermal contributions to the EOS. These differences are physically consistent with the influence of different types of interactions between 
atoms in the two materials. BN has slightly higher ionic character than C due to the differences between the electronegativity of B and N, associated with
dipolar interactions between the non-identical atoms.

\subsection{Zero-point motion effects\label{subsec:zpm}}
\begin{figure}
\centering\includegraphics[width=0.48\textwidth]{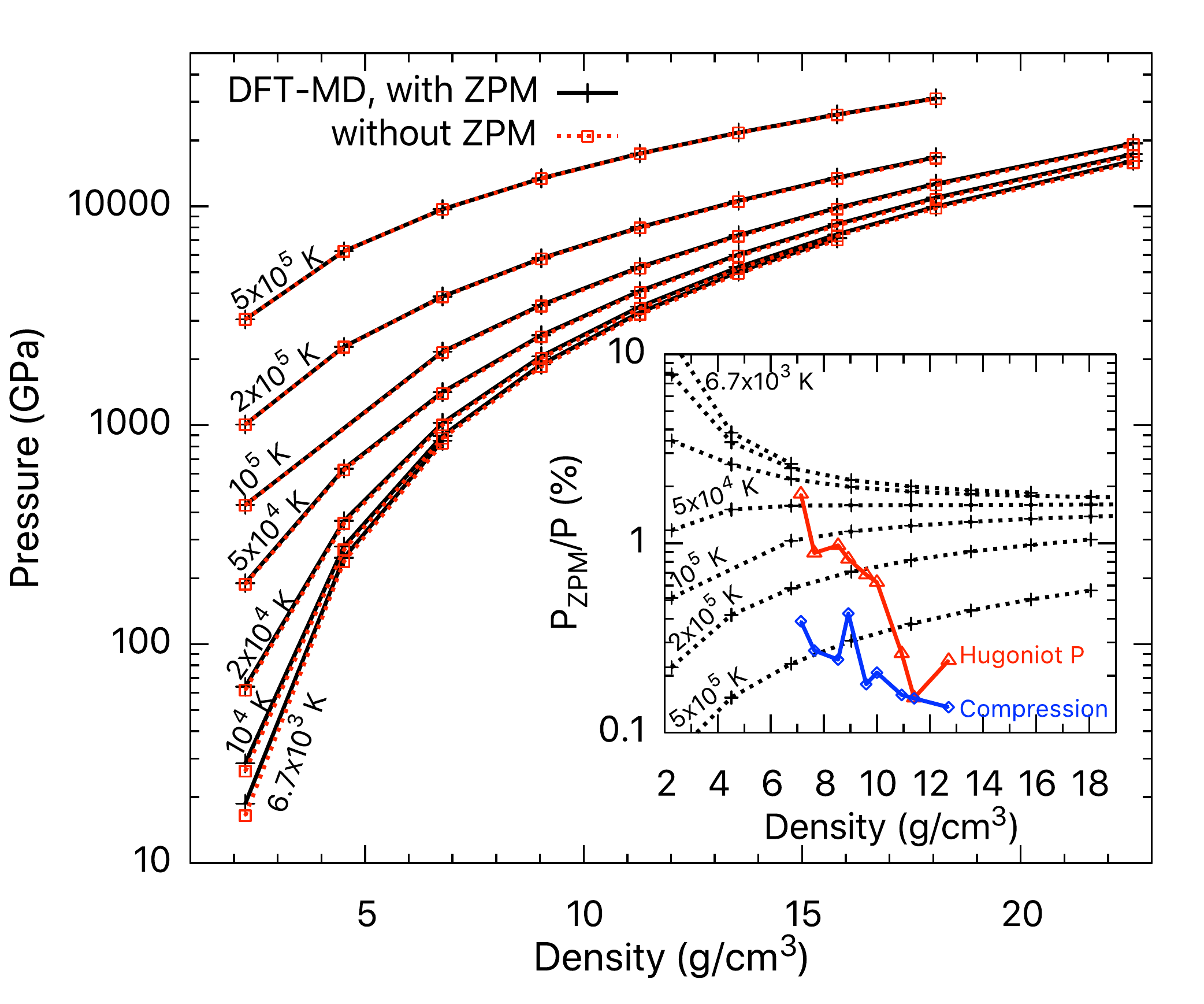}
\caption{\label{fig:zpm} Zero-point motion effects on the pressure of BN as a function of density along several isotherms. The inset shows the percent increase in pressure for the EOS (black) and along the Hugoniot (red) and percent decrease in compression ratio along the Hugoniot (blue).}
\end{figure}

We have also examined the effect of Zero-point motion (ZPM) on the EOS and Hugoniot of BN. In order to do this, we implement the Debye model~\cite{Wallace2002} to estimate the magnitude of the EOS contributions due to ZPM. This correction reasonably account for the nuclear quantum effects that have been neglected in the our Born-Oppenheimer MD simulations. According to the Debye model, the harmonic vibration energy can be approximated by $\delta E=9k_B\Theta_D(V)/8$, where $\Theta_D(V)$ is
the volume-dependent Debye temperature and is related to the ambient-density via $\Theta_D(V)=\Theta_D(V_0)(\rho/\rho_0)^\gamma$ with $\gamma$ being the Gr\"{u}neisen parameter, and the corresponding
pressure $\delta P=9\gamma k_B\Theta_D(V)/8V$.
We take the values $\Theta_D(V_0)=1900$ K and $\gamma=1.1$ for c-BN from previous measurements and calculations~\cite{WANG2017276jpcs,deKoker2012jpcm}, apply the corrections to our EOS data from DFT-MD (pwPAW) and evaluate the changes in the Hugoniot curve. The results are summarized in Fig.~\ref{fig:zpm}.

Our results show that ZPM causes a pressure increase by over 10\% at 6.7$\times10^3$~K and ambient density. This percentage difference decreases gradually to $\sim1\%$ at 20~g/cm$^3$. The differences dramatically decrease as temperature becomes higher, more so at lower densities.
The effect of ZPM on Hugoniot, however, is small. For example, the compression ratio decreases by up to 0.01 (0.4\%) for the temperature range 6.7$\times10^3$--5.1$\times10^5$ K considered in our DFT-MD (pwPAW) simulations. This is similar to what we have seen in carbon-hydrogen systems~\cite{Zhang2018}. These findings indicate that the ZPM should be carefully addressed when studying the the low-$Z$ materials in the condensed matter regime, but is negligible for studying the shock Hugoniot of them in the high-energy-density plasma state. 

\section{Conclusions}\label{sec:conclusion}
In this work, we present a comprehensive study of the EOS of BN over a wide range of pressures and temperatures by implementing several computational methods, including PIMC, DFT-MD using standard plane-wave basis and PAW or ONCV potentials, ACTEX, FOE, SQ, {\footnotesize MECCA}, and TF. We use the PIMC, DFT-MD, and {\footnotesize ACTEX} data to construct two new EOS tables (X2152 and X2151) for BN using the QEOS model. 

Our EOS data by PIMC, FOE, SQ, and {\footnotesize MECCA} show good consistency at $10^6$ K where 1s electrons are ionized.
Our findings establish SQ as an accurate method capable of calculating the EOS with high precision and accuracy comparable to PIMC.
Our detailed EOS comparison provides strong evidences that
cross validate both the PIMC and the DFT-MD approaches for EOS studies of the partially ionized, warm-dense plasmas.

At 2.5--3.2$\times10^6$~K and 1.0--1.3$\times10^5$~GPa, our PIMC, {\footnotesize ACTEX}, and {\footnotesize MECCA} calculations uniformly predict a maximum compression of $\sim$4.55 along the shock Hugoniot for h-BN ($\rho_\text{i}$=2.26 g/cm$^3$), which originates from $K$ shell ionization. This compression is underestimated by TF models by $\sim$0.2. The maximum compression decreases to 4.47 for c-BN ($\rho_\text{i}$=3.45 g/cm$^3$) and increases to 4.59 for $\rho_\text{i}$=2.15 g/cm$^3$.

We also report Hugoniot data up to $\sim 2650$ GPa from experiments at the Omega laser facility. The measured data show good agreement with our theoretical predictions based on DFT-MD.

By comparing QEOS models with the electron thermal term constructed in different ways (Purgatorio, TF, or hybrid), we find that the shock Hugoniot can be well reproduced by fitting the QEOS models to the pressures in the EOS calculated from first principles.  Consistent with our previous studies, we find that the Purgatorio-based EOS models provide the
best agreement with both internal energies and pressures from first principles calculations.  Because the largest differences in the Hugoniot response of the models occurs 
near peak compression, performing experiments for materials near peak compression~\cite{Swift2012,Kritcher2016,Nilsen2016,Swift2018,Doppner2018} would provide a rigorous experimental test of our understanding 
of electronic structure in high energy density plasmas.  It would also be worthwhile to pursue experiments that provide measurements of the temperature and the pressure in either 
Hugoniot or off-Hugoniot experiments, which would provide data to validate the first principle 
calculations.

We find the shock Hugoniot profiles of isoelectronic materials BN and C are very similar, with the compression peak of C being slightly sharper. This is explained by the differences between the 1s level of C and those of B and N.  Based on the similarities of these materials in the laser-induced shock regime, BN ablators would be expected to behave similarly to 
HDC ablators.  While the impact of the condensed phase microstructure of the materials may also be an important consideration in the compressive, ICF 
regime where much of the ablator is still present during the implosion phase, the microstructure should be less consequential to the behavior of exploding pushers where most of the 
ablator has been vaporized.


\section{Appendix}
\subsection{\label{subsec:ONCV}Optimized norm-conserving Vanderbilt pseudopotentials}

\begin{figure}[htb]
\centering
\includegraphics[width=0.48\textwidth]{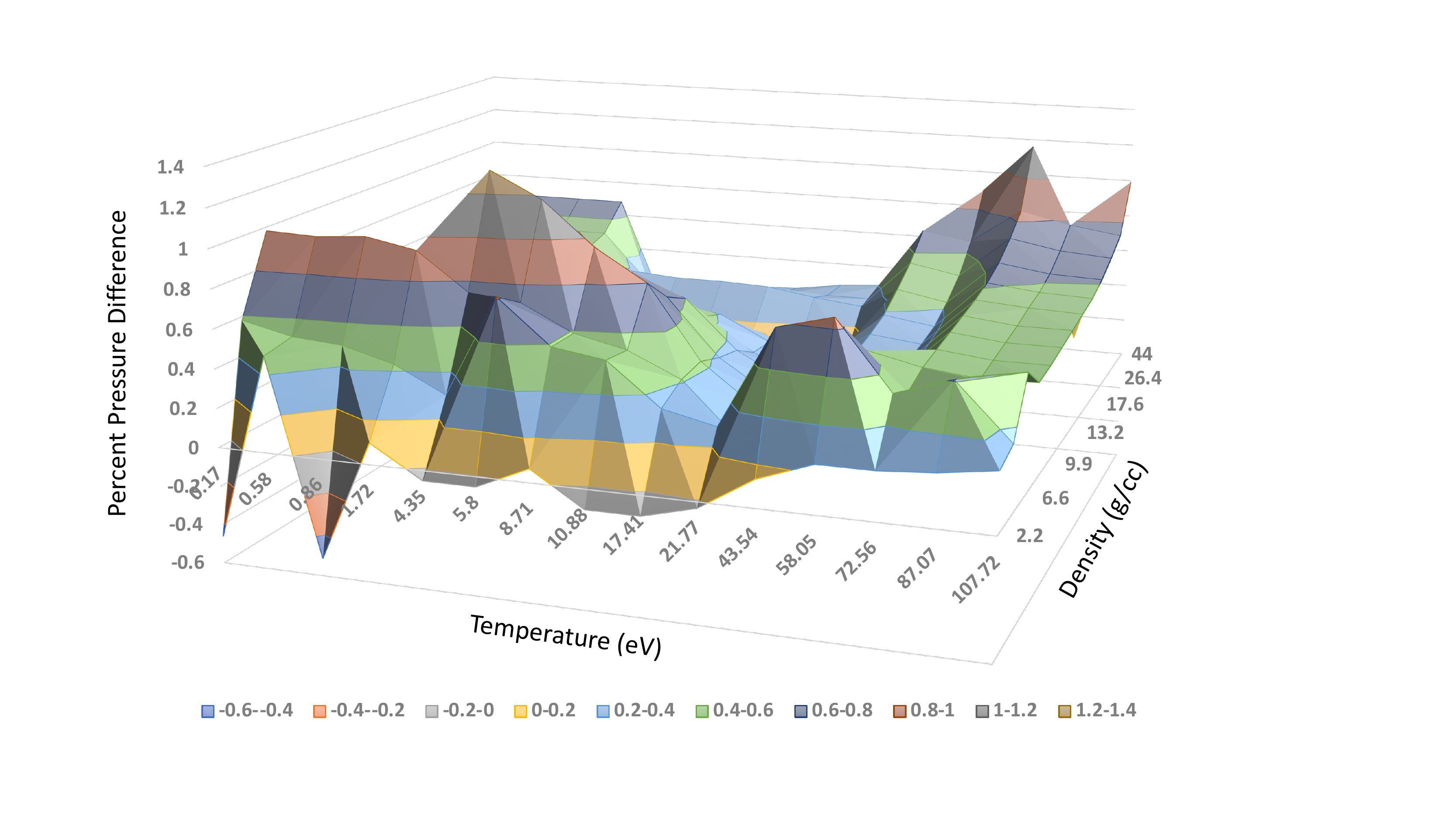}
\caption{Percent pressure difference between calculations using ONCV all-electron pseudopotentials and regularized Coulomb potentials for BN  in the 
cubic phase. For most of the phase points examined in this study, the difference is within 1\% except a few cases where the difference is slightly greater.}
\label{fig:oncv2.07-Col}
\end{figure}

We employed ONCV pseudopotentials~\cite{oncv13} for a subset of DFT-MD calculations, in addition to the FOE and SQ calculations. Fully nonlocal two-projector norm-conserving pseudopotentials were generated.  The resulting potentials have an accuracy in electronic structure properties comparable to {\footnotesize VASP} PAW and all-electron calculations.  Due to the wide range of density and temperature grids used in the EOS table generation, we have constructed two versions of ONCV pseudopotentials for B and N to reduce projector overlap and core-state ionization under these extreme conditions.  The first set of ONCV pseudopotentials have $2s^{2}$ and $2p^{1}$ valence states for B and $2s^{2}$ and $2p^{3}$ valence states for N, respectively. The second set of ONCV pseudoptentials are all-electron pseudopotentials that include $1s^{2}$ valence. The parameters associated with the corresponding psuedopotentials are listed in Table~\ref{tab:oncv}. To cross check the accuracy of the ONCV pseudopotentials we compared  calculated pressures with regularized Coulomb potentials ($r_{c}=0.02$~Bohr and kinetic-energy cutoff of 6000~Ha) for solid c-BN 
phase at each density-temperature point in the DFT-MD simulations. The overall agreement between ONCV pseudopotentials and regularized Coulomb potentials is within 1\% except a few points slightly greater. As an example, Figure~\ref{fig:oncv2.07-Col} shows the percent difference of pressure between all-electron ONCV pseudoptentials and Coulomb potentials for 
c-BN within the density-temperature grid employed in the DFT-MD simulations. The pressure difference ranges from $-0.6\%$ to $1.4\%$, with the larger differences 
in the low-temperature, low-density regions. 

\begin{table}
\centering
\begin{tabular}{ccccccc}
\hline 
Species & Valence & $r_{c}$ & $K_\text{Cutoff}$ & Note \\ 
 &  &  (Bohr) & (Ha)  & \\ 
\hline 
B & $2s^{2}2p^{1}$ & 1.125 & 35  & pwONCV \\ 
B & $1s^{2}2s^{2}2p^{1}$ & 0.6 & 160 & FOE \\ 
B & $1s^{2}2s^{2}2p^{1}$ & 0.6 & -- & SQ \\ 
N & $2s^{2}2p^{3}$ & 1.2 & 35 & pwONCV \\ 
N & $1s^{2}2s^{2}2p^{3}$ & 0.65 & 160 & FOE \\
N & $1s^{2}2s^{2}2p^{3}$ & 0.65 & -- & SQ \\
\hline 
\end{tabular} 
\caption{Parameters used to generate ONCV psuedopotentials for B and N. Bulk properties calculated from these pseudopotentials were benchmark against {\footnotesize VASP} PAWs and regularized Coulomb potentials. 
$r_c$ and $K_\text{Cutoff}$ denote the local potential core radius and the kinetic energy cutoff, respectively.
The potentials for SQ are similar to those in FOE, but used higher continuity at $r_c$ to remove cusps and improve convergence.}
\label{tab:oncv}
\end{table}

\subsection{Mean-field Thomas-Fermi and average-atom in jellium (Purgatorio)}\label{TFPurg}

Our EOS models are developed on a broad grid in phase space, spanning many decades in both temperature and pressure.  As such, we require efficient 
methods for computing the electron thermal contribution to the EOS.  In this work, we apply two methods for this purpose, both of which are based on 
density functional theory.  Our TF calculations are based on the generalized  theory of~\citet{FMT:49}.  In contrast to the TF approach, which assumes a uniform Fermi distribution of states and thus does not explicitly include discretized states, 
Purgatorio solves the electronic structure problem for an atom-in-jellium within LDA self-consistently, and thus allows for the inclusion of discretized states.\cite{Purgatorio2006,Sterne2007}  

For computing the EOS of 
mixtures, such as BN, from either Purgatorio or TF, we apply a constant electron pressure mixing rule, following the prescription outlined in
Ref.~\onlinecite{JN:14}. Briefly, if $x_1$ and $x_2$ represent concentrations
of the two ions, then the Wigner-Seitz (WS) volume per ion of the
plasma is required to be the weighted sum of the WS volumes of its two constituent ions:
\begin{equation}
\frac{x_1 A_1+x_2 A_2}{N_A \rho} = x_1\frac{A_1}{N_A \rho_1} + x_2\frac{A_2}{N_A \rho_2}. \label{one}
\end{equation}
In the above, $\rho$, $\rho_1$ and $\rho_2$ are the densities of the plasma and its ionic components,
$A_1$ and $A_2$ are atomic weights of the constituent ions and $N_A$ is the Avagadro constant.
This equation is supplemented by the requirement that the free electron density of the
plasma be unique:
\begin{equation}
p_e(1) = p_e(2).  \label{two}
\end{equation}
Moreover, since the pressure in the TF theory depends only on $T$ and $\mu$, it follows that
the electron density in the plasma is also unique $n_e(1) = n_e(2)$.
In the TF method, the free electron density $n_e(i)$ associated with ion $i$ is determined by solving the TF equations
for the ion at specified values of temperature $T$ and density $\rho_i$. 
 At a given value of $T$, Eqs.~\ref{one}-\ref{two}  provide
two equations that can be solved to give values of the unknown densities $\rho_1$ and $\rho_2$.
Inasmuch as $n_e(i)$
is a monotonic function of $\mu_i$, it follows that the chemical potential
is also unique $\mu_1 = \mu_2$.

To create an EOS table for two-ion plasmas, we first choose a $T$ grid uniformly spaced
on a logarithmic scale. For each temperature on the $T$ grid, we solve the TF equations for
the two ions on density sub-grids ranging from 1/2 to 5 times the respective cold-matter densities.
The properties of ion 2: $\rho_2$, $p_2$, and $\mu_2$,  
considered as functions of electron density $n_e(2)$ are interpolated onto the electron density grid of ion 1.  
In this way, Eq.~\ref{two} is automatically satisfied at each point on the $n_e(1)$ grid. We can verify that 
this procedure leads to $p = p_2 = p_1$ and $\mu = \mu_2=\mu_1$ for the interpolated values. Furthermore, 
we can now determine the density $\rho$ of the two-ion plasma at each point on the $n_e(1)$ grid  
using Eq.~\ref{one}. In this way, an EOS table is created for $p$ as a function of $\rho$ and  $T$.
The approach is similar for a Purgatorio-based EOS table for a multi-component material: we perform Purgatorio calculations for the individual elements on a ($\rho$, $T$) grid and
mix the tables according to the pressure equality denoted in Eq.~\ref{two}.

\begin{acknowledgments}  
This work was in part performed under the auspices of the U.S. Department of Energy by Lawrence Livermore National Laboratory under Contract No. DE-AC52-07NA27344.
Computational support was provided by LLNL high-performance computing facility (Quartz and Jade) and the
  Blue Waters sustained-petascale computing project 
  (NSF ACI 1640776).
  Blue Waters is a joint effort of the University of Illinois at 
  Urbana-Champaign and its National Center for Supercomputing Applications. 
B.M. is supported by the U. S. Department of Energy (grant DE-SC0016248) and the University of California. S.Z. is partially supported by the PLS-Postdoctoral Grant of LLNL.
D.D.J. and A.V.S. were partially funded for KKR results by the U. S. Department of Energy, Office of Science, Fusion Energy Sciences through Ames Laboratory, which is operated by Iowa State University for the U.S. DOE under contract DE-AC02-07CH11358. 
J.P. thanks D.R. Hamann for helpful discussions and code facilitating the construction of robust all-electron ONCV potentials. 
We appreciate Zsolt Jenei for Raman spectroscopy
and John Klepeis, Tadashi Ogitsu and John Castor for useful discussion.

This document was prepared as an account of work sponsored by an agency of the United States government. Neither the United States government nor Lawrence Livermore National Security, LLC, nor any of their employees makes any warranty, expressed or implied, or assumes any legal liability or responsibility for the accuracy, completeness, or usefulness of any information, apparatus, product, or process disclosed, or represents that its use would not infringe privately owned rights. Reference herein to any specific commercial product, process, or service by trade name, trademark, manufacturer, or otherwise does not necessarily constitute or imply its endorsement, recommendation, or favoring by the United States government or Lawrence Livermore National Security, LLC. The views and opinions of authors expressed herein do not necessarily state or reflect those of the United States government or Lawrence Livermore National Security, LLC, and shall not be used for advertising or product endorsement purposes.
\end{acknowledgments}

\end{document}